\documentclass[twocolumn,10pt]{IEEEtran}
\pdfoutput=1
\usepackage{comment}
\includecomment{toexclude}
\usepackage{amsmath,amsfonts,amssymb}
\usepackage{cite}
\usepackage{verbatim}
\usepackage{bm}
\usepackage{algorithm}
\usepackage{graphicx}
\usepackage{xcolor}
\usepackage{url}

\newcommand{\figsize}{0.9\columnwidth}

\newcommand{\yf}{{\mathbf{y}}}

\newcommand{\Yf}{\mathbf{Y}}

\newcommand{\Xt}{\mathbf{X}}
\newcommand{\xf}{{\mathbf{x}}}

\newcommand{\Hf}{{\mathbf{H}}}
\newcommand{\hct}{\mathbf{h}}

\newcommand{\Ht}{\mathbf{H}}

\newcommand{\F}{\mathbf{F}}
\newcommand{\A}{\mathbf{A}}
\newcommand{\av}{\mathbf{a}}

\newcommand{\D}{\mathbf{D}}

\newcommand{\B}{\mathbf{B}}
\newcommand{\Pc}{\mathbf{P}}
\newcommand{\p}{\mathbf{p}}

\newcommand{\I}{\mathbf{I}}

\newcommand{\W}{\mathbf{W}}

\newcommand{\diag}{\text{diag}}

\newcommand{\C}{\mathbf{C}}
\newcommand{\cv}{\mathbf{c}}

\newcommand{\U}{\mathbf{U}}
\newcommand{\V}{\mathbf{V}}

\title{Framework for a Perceptive Mobile Network using Joint Communication and Radar Sensing}

\author{{Md. Lushanur Rahman, J. Andrew Zhang,~\IEEEmembership{Senior~Member,~IEEE}, Xiaojing Huang,~\IEEEmembership{Senior~Member,~IEEE}, \\
		Y. Jay Guo,~\IEEEmembership{Fellow,~IEEE}, and Robert~W.~Heath~Jr,~\IEEEmembership{Fellow,~IEEE}} 
\thanks{Md. Lushanur Rahman, J. Andrew Zhang, Xiaojing Huang and Y. Jay Guo are with University of Technology Sydney, (UTS), Global Big Data Technologies Centre (GBDTC), Australia. Email: MdLushanur.Rahman@student.uts.edu.au; \{Andrew.Zhang; Xiaojing.Huang; Jay.Guo\}@uts.edu.au}
\thanks{Robert W. Heath Jr is with The University of Texas at Austin, Austin, TX 78712, USA. Email: rheath@utexas.edu.}
\thanks{The results in this paper were partially presented in VTC Spring 2017 \cite{Zhang17basic} and ICEAA 2017 \cite{Zhang17strip}.} 
}

\begin{document}

\maketitle \thispagestyle{empty} \pagestyle{empty}

\begin{abstract}
 In this paper, we develop a framework for a novel perceptive mobile/cellular network that integrates radar sensing function into the mobile communication network. We propose a unified system platform that enables downlink and uplink sensing, sharing the same transmitted signals with communications.  {We aim to tackle the fundamental sensing parameter estimation problem in perceptive mobile networks, by addressing two key challenges associated with sophisticated mobile signals and rich multipath in mobile networks. To extract sensing parameters from orthogonal frequency division multiple access (OFDMA) and spatial division multiple access (SDMA) communication signals, we propose two approaches to formulate it to problems that can be solved by compressive sensing techniques. Most sensing algorithms have limits on the number of multipath signals for their inputs. To reduce the multipath signals, as well as removing unwanted clutter signals, we propose a background subtraction method based on simple recursive computation, and provide a closed-form expression for performance characterization.} The effectiveness of these methods is validated in simulations.
\end{abstract}


\section{Introduction}\label{sec:intro}

{
The joint communication and radar sensing (JCAS, aka Radar-Communication) technology is receiving increasing interests thanks to its capability in integrating communication and radar sensing into one system, sharing the same transmitted signals and a majority of hardware and signal processing modules \cite{RN32,Kumari17,7782415,RN15}. One major potential application for the JCAS technology is in vehicular networks \cite{Kumari17 , 8057284}, where communication signals can also be used for sensing the environment for object detection and collision avoidance. Another potentially significant application is in mobile (aka cellular) networks. Having the largest broadband coverage and powerful infrastructure, JCAS-enabled mobile networks can potentially become a ubiquitous radio sensor, while providing simultaneous communication service.  

JCAS-enabled mobile network can be significantly different to passive bistatic and multistatic radar systems which use mobile communication signals for sensing \cite{Hack14,Gogineni14, Abdullah2016}. In the JCAS network, receivers know the detailed structure of the transmitted signal, such as resource allocation for time, frequency and space, and the transmitted data symbols (either directly known or through demodulation). Such knowledge on signal structure is important for coherent detection, which enables accurate estimation for sensing parameters. Comparatively, most passive radar sensing can only use non-coherent detection, where typically only the power, angles and Doppler information can be extracted  from the received signals at degraded performance \cite{Hack14,Gogineni14}. In a mobile network environment, without the knowledge of the signal structure, passive sensing also lacks the capability of interference suppression, and cannot separate multi-user signals from different transmitters (signal sources).

There have been limited JCAS results closely related to modern mobile networks. In \cite{Braun14}, some early work on using orthogonal frequency-division multiplexing (OFDM) signal for sensing was reported. { In \cite{article}, sparse array optimization was studied for multiple-input multiple-output (MIMO) JCAS systems. In \cite{Rong17}, the multiple access performance bound is derived for a multiple antenna JCAS system.} In \cite{Liu17ada}, mutual information for an OFDM JCAS system is studied, and power allocation for subcarriers is investigated based on maximizing the weighted sum of the mutual information for radar and communications.
In \cite{Liu18}, waveform optimization is studied for minimizing the difference between the generated signal and the desired sensing waveform under the constraints of signal-to-interference-and-noise ratio for multiuser MIMO (aka spatial division multiple access, SDMA) downlink communication. A multi-objective function is further applied to trade off the similarity of the generated waveform to the one desired for communication and sensing \cite{Liu18mumimo}. These studies involve some key signal formats in modern mobile networks, such as MIMO, multiuser MIMO, and OFDM. However, there is very limited work on how JCAS can actually be realized at a system level in the mobile network, and how radar sensing can be done based on modern mobile communication signals, which is a fundamental and challenging problem.  

In this paper, we develop a framework for integrating radar sensing into current communication-only mobile network using JCAS technologies, by synthesizing and extending our earlier work in \cite{Zhang17basic, Zhang17strip}. We call it \textit{perceptive mobile network}. This framework includes both a \textit{system platform} for unified radar sensing and novel \textit{sensing solutions}. We set up the perceptive mobile network on a system platform with key components and technologies in modern mobile networks, such as antenna array, broadband through e.g. channel aggregation, multi-user MIMO and orthogonal frequency-division multiple access (OFDMA). The system platform provides system-level integration for communication and radar sensing, and unifies three types of sensing based on the communication signals. The sensing solutions address critical challenges in estimating \textit{sensing parameters} including time delay, angle-of-arrival (AoA), angle-of-departure(AoD), Doppler shift and magnitude of multipath signals. These challenges are caused by both sophisticated signal format and massive multipath signals due to complicated signal propagation environment.

The first challenge for sensing parameter extraction in perceptive mobile networks is due to the sophisticated signal structure. The communication signals, which are also used for sensing, can be randomly modulated with multiple users’ symbols using multiuser-MIMO and OFDMA technologies, and be fragmented for each user - discontinuous over time, frequency or space. This will be detailed in Section \ref{sec:formation}. Such signal structure makes most existing sensing parameter estimation techniques not directly applicable. For example, active radar sensing technologies mainly deal with optimized or unmodulated transmitted signals \cite{Haim08,Liang11}; most passive bistatic and multistatic radars consider simple single carrier and OFDM signals \cite{Hack14,Abdullah2016, Gogineni14, 7833233};  and channel estimation techniques developed for modern mobile networks mainly focus on estimating channel coefficients instead of detailed channel compositions represented by the sensing parameters. In addition, conventional spectrum analysis and array signal processing techniques such as MUSIC and ESPRIT require continuous observations, which are not always available here. Therefore, new sensing techniques need to be developed for estimating sensing parameters from the complicated and fragmented signals. We will show that compressive sensing (CS) is an excellent candidate technology for this problem, after proper signal formulation.

The second challenge for sensing parameter estimation comes from the rich multipath in mobile networks. Most sensing parameter estimation algorithms can only process signals containing a limited number of multipath signals. Hence we need a preprocessing method which can divide signals into different groups where each group has significantly reduced number of multipath. At a minimum, it is essential to separate and reduce non-information-bearing multipath signals from the input. Such unwanted multipath is called as \textit{clutter} in the radar literature. Typical radar systems are optimized for sensing a limited number of objects in open spaces using narrow beamforming, and clutter has notably different features from useful reflections returned from ground, sea, rain etc. \cite{RN32,RN54}. Most known algorithms in radar systems, such as subspace method \cite{RN52}, CLEAN algorithm \cite{6020778}, independent component \cite{RN50} are adapted to such scenarios. In the perceptive mobile network, we define \textit{clutter} as unwanted multipath signals that contain little new information. The clutter hence is mainly referred to the multipath signals from permanent or long-period static objects when both the transmitter and receiver of the sensing devices are static. Due to the different signal propagation environment, suppression requirements and applicable sensing algorithms,  existing clutter suppression techniques developed for radar systems, e.g., those in \cite{RN54, RN50, RN52}, may not directly render the clutter reduction here. Most of them are also applied after sensing algorithms, and hence cannot achieve the goal of reducing multipath input to the sensing algorithms.
}

{
In this paper, we focus on studying system-level integration of sensing function into mobile communication networks, and investigating how to address the two critical challenges for sensing parameter estimation as described above. As an initial piece of work in this new domain, our proposed algorithms here mainly intend to demonstrate the feasibility and methodology, but are yet to be optimized for complexity and performance. Our major contributions in this paper are as follows:
\begin {itemize}
\item We introduce a unified system platform that enables three types of sensing to be integrated with mobile communications. We present the required changes for hardware and system in existing mobile networks. We also provide signal formulation for the three types of sensing, and show that they can be represented by a common expression, which enables the application of common sensing algorithms.
\item We present two schemes for estimating sensing parameters from sophisticated communication signals with modulations of OFDMA and multiuser-MIMO. The first one is \textit{direct estimation} that uses the received mobile signals directly as inputs to sensing algorithms, assuming that the transmitted information data symbols are known. The second is \textit{indirect estimation} based on \textit{signal stripping}. It simplifies the signal input to sensing algorithms by removing (demodulated) data symbols and decorrelating users using conventionally estimated channels in communications. Upon the formulated signal models based on these two methods, we demonstrate how sensing parameters can be estimated via employing one-dimensional (1D) compressive sensing (CS) algorithms. The proposed 1D CS algorithms are particularly useful when there is only sufficient  measurements in one dimension, which could be typical in current and near-future systems.
\item We propose a low-complexity \textit{background subtraction} method for reducing clutter from the input to sensing algorithms. It reconstructs clutter using simple recursive computation, and allows separation of signals with largely separated Doppler frequencies. We also provide closed-form expressions to show how the reconstruction performance and noise are related to the parameters in the recursion equation. This method is not only capable of removing clutter, but also has the potential of dividing multipath signals into different groups according to their Doppler shift values.
\end{itemize}}

The rest of this paper is organized as follows: In Section \ref{sec-problem}, we introduce the system platform for the perceptive mobile network. In Section \ref{sec:formation}, we provide mathematical models for the sensing problems. In Sections \ref{sec-onlysymbol} and \ref{sec-channelsymbol}, the direct and indirect sensing schemes are presented, respectively. Section \ref{sec-clutter} presents the background subtraction method for clutter suppression. In Section \ref{sec-simu}, simulation results are provided to validate the effectiveness of the proposed framework and sensing algorithms. Section \ref{sec-conclusion} concludes the paper.
  
Notations: $(\cdot)^H$, $(\cdot)^T$ and $(\cdot)^c$ denote the Hermitian transpose, transpose and conjugate of a matrix/vector, respectively. $|\cdots|$ denotes the element-wise absolute value, $(\A)_{n,m}$ denotes the $(n,m)$-th element of the matrix $\A$, $(\A)_{\cdot,m}$ and $(\A)_{m,\cdot}$denotes the m-th column and row of $\A$, respectively, $\{a_n\}$ denotes a vector with elements $a_n$, $\diag\{a_n\}$ denotes a diagonal matrix with diagonal elements $a_n$.

\section{System Platform for the Perceptive Mobile Network }\label{sec-problem}

{ Our proposed system platform  aligns with the specification of the evolution of mobile networks, such as 5G. In this section, we describe the system model, the supported sensing operations and the required modifications to existing mobile communication infrastructure. }

\begin{figure}[t]
 \centering
 \includegraphics[width=0.9\columnwidth]{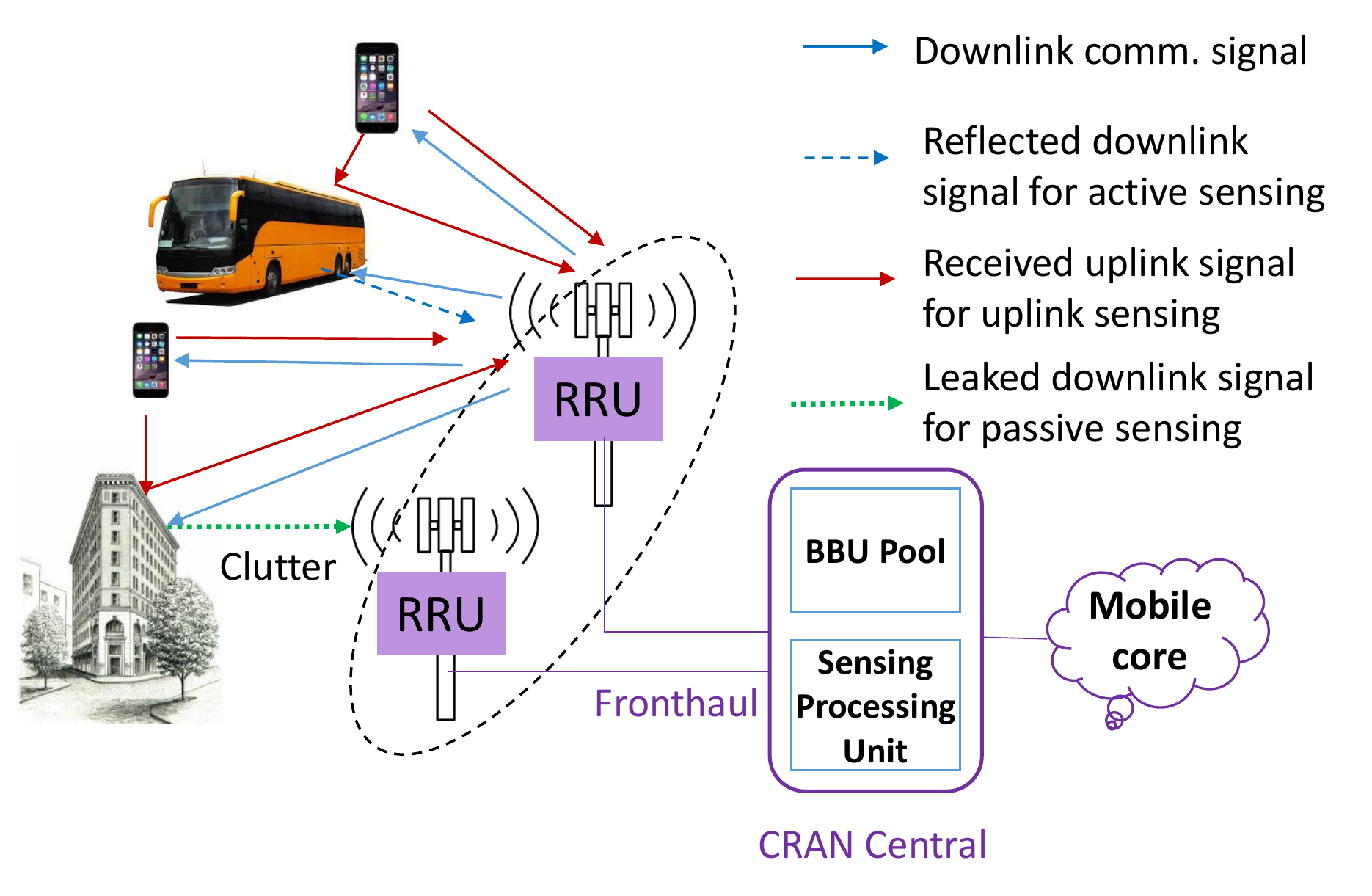}
 \caption{Proposed System Model.}
 \label{fig-system}
 \end{figure}

{\subsection{System Model}\label{sec-platform}
We assume a cloud-radio-access network (CRAN) architecture using multiuser-MIMO and OFDMA technologies. Fig.\ref{fig-system} shows the  CRAN architecture based system model of the proposed perceptive mobile network. In this model, cooperative remote radio units (RRU), are densely distributed and synchronized in clock. Signal processing for both cellular communication and radio sensing based on collected signals from these RRUs is done centrally in CRAN central, which includes the baseband unit (BBU) pool for communication and the sensing processing unit. We assume that cooperative RRUs are within the signal coverage area of each other. 
All RRUs' clocks are synchronized, typically via GPS. A typical communication scenario is as follows: several RRUs work cooperatively to provide connections to mobile stations (MSs), using multiuser MIMO techniques over the same subcarriers. While we consider CRAN it could work for a standalone base-station (BS) too. So hereafter we will use CRAN central and BS without differentiating them.}

We focus on the case where radio sensing is conducted in the BS, although MS-side sensing is also possible. Compared to MS, BS has advantages of networked connection, flexible cooperation, large antenna array, powerful computation capability, and known and fixed locations.

\subsection{Supported Sensing Operations}\label{sec-updown}
In the perceptive mobile network, the transmitted signal from BSs or mobile stations (MSs) is used for both communication and sensing. The signal may be optimized jointly for the two functions, and one example is available from \cite{Liu18}. We define \textit{uplink and downlink sensing}, to be consistent with uplink and downlink communications.  In uplink sensing, the used sensing signal is from MSs. In downlink sensing, the sensing signals are from BSs.  The downlink sensing is further classified as \textit{Downlink Active Sensing} and \textit{Downlink Passive Sensing}, for the cases when a RRU collects the echoes from its own and from other RRUs transmitted signals, respectively.

It is important to note that in a distributed antenna system such as CRAN, sensing is for the environment surrounding a specific transmitter and receiver, and hence it is separately done for each node (RRU in CRAN), although some joint processing is possible. 
 
\subsubsection{Downlink Active Sensing}
We refer to \textit{downlink active sensing} as the case that a RRU uses reflected downlink communication signals from its own transmitted signal for sensing. In this case, similar to a mono-static radar, transmitter and receiver are co-located although they may have two independent antennas separated in space. This will enable a RRU to sense its surrounding environment. 

\subsubsection{Downlink Passive Sensing}

\textit{Passive sensing} is typically referred to the case when a third receiver outside the communication system exploits the communication signal for sensing. Here we use downlink passive sensing for the case where a RRU uses the downlink communication signals received from other RRUs for sensing. Depending on the distance between RRUs, reflected signals from other RRUs or the RRU itself may arrive at different time segment or overlapped. Downlink passive sensing senses the environment among RRUs.

\subsubsection{Uplink Sensing}
BS uses the uplink communication signal from MS transmitters for \textit{uplink sensing}. Uplink sensing estimates relative, instead of absolute, time delay parameters because the timing in MS transmitters and RRU receivers is typically not aligned. This timing ambiguity may be removed by using techniques developed from the triangulation techniques in localization. Uplink sensing senses MSs and the environment between MSs and RRUs.

{
Downlink sensing can potentially lead to more accurate sensing results than uplink sensing because RRUs exhibit more advanced transmitter capability than MS and also the transmitted data symbols are centrally known in the downlink. In addition, privacy issue is almost not a problem in downlink sensing because the sensed results are not directly linked to any MSs.

The signals available for sensing can come from several sources, such as the demodulation reference signal (DMRS) and the whole data payload signals in 5G New Radio. The transmitted signals can also be optimized jointly for both communication and radar sensing, such as those proposed in \cite{Liu17ada,Liu18, Liu18mumimo}. In this paper, we focus on using the always-available whole data payload, as DMRS signals are random and may be insufficient for high-resolution sensing.
}
 
\subsection{Required System Modification}\label{sec-structure}
{We now describe potentially required modifications on hardware and system, in order to evolve current communication only mobile networks to perceptive mobile networks. 

Uplink sensing can be implemented without requiring changes on hardware and system architectures of current mobile systems, in the presence of the timing ambiguity problem. Alternatively, dedicated (static) MSs that are clock-synchronized to BSs can be used, which would be the most convenient way for achieving non-ambiguity sensing in the perceptive mobile networks.

On the other hand, downlink sensing requires changes to hardware, and the extent of changes depends on the network duplexing mode. Basically, downlink sensing requires a transceiver to work on the\textit{ full duplex} mode, where receiver and transmitter need to operate at the same time. This causes transmitted signal leakage, which can easily overwhelm the reflected echoes for downlink sensing  without modifying the current hardware. The full duplex technology, which typically uses antenna separation, RF suppression and baseband suppression to mitigate the leakage, is a promising enabling technology in a long term \cite{7105651}. Although it progresses well, full duplex MIMO is still challenging to realize in practice due to antenna cross-talk and coupling. 

Two near-term solutions to downlink sensing are as follows:
\begin{itemize}
	\item Using two sets of spatially well-separated antennas for transmitting and receiving. Nevertheless, this requires extra antenna installation space and can increase the overall cost.
\item Deploying RRUs that only work on the receiving mode. They can be configured as working in the sensing mode only or in both communication and sensing modes. 
\end{itemize}
To implement these near-term solutions, changes to the hardware are required. For time-division-duplexing (TDD) systems, the change is minor since a TDD transceiver generally uses a switch to control the connection of antennas to the transmitter or receiver.  For frequency division duplexing (FDD) systems, the receivers may not be capable of working on downlink frequency bands. From this point of view, it is more cost-effective to implement downlink sensing in TDD than in FDD systems. 

In addition to the essentially required system modifications as described above, there are also many research challenges and opportunities for joint system design and optimization, such as joint waveform optimization \cite{Liu17ada,Liu18, Liu18mumimo} and joint antenna placement and sparsity optimization \cite{Wang18}. Such joint design and optimization, on top of solutions to fundamental problems for the perceptive mobile networks such as these being discussed in this paper, can be expected to improve the performance of integrated systems significantly.
}
\section{Formulation of Sensing Parameter Estimation}
\label{sec:formation}
{
In this section, we formulate the signal models used for estimating sensing parameters. We first introduce general system and channel models, and extend them to downlink and uplink sensing,  and then provide a generalized on-grid model by quantizing the delay. We show that downlink and uplink can be represented by a common model, which enables common sensing algorithms. We also provide detailed justification for the choice of this on-grid model and the corresponding compressive sensing techniques. We then present two schemes for sensing parameter estimation in Section \ref{sec-onlysymbol} and \ref{sec-channelsymbol} based on the formulated models here.} 

\subsection{General System and Channel Models}

We consider a CRAN system with $Q$ RRUs and each RRU has a uniform linear array (ULA) with $M$ antenna elements and antenna interval of half wavelength. These RRUs cooperate and provide links to $K$ users {through multiuser-MIMO and OFDMA technologies, i.e., each user may occupy and share only part of the total subcarriers with other users through multiuser-MIMO.} Each user has a ULA of $M_T$ elements. For both uplink and downlink, we assume that data symbols are first spatially precoded, and an IFFT is then applied to each spatial stream. The time domain signals are then assigned to the corresponding RRUs. Let $N$ denote the number of total subcarriers and $B$ the total bandwidth. Then the subcarrier interval is $f_0=B/N$ and OFDM symbol period is $T_s=N/B+T_p$ where $T_p$ is the period of cyclic prefix. 

Assume a planar wave-front in signal propagation. The array response vector of a size-$M$ ULA is given by
\begin{align}
\av(M,\theta)=[1,e^{j\pi \sin(\theta)},\cdots, e^{j\pi (M-1)\sin(\theta)}]^H,
\end{align}
where $\theta$ is either AoD or AoA.

Let the AoD and AoA of a multipath be $\theta_{\ell}$ and $\phi_{\ell}$, $\ell\in[1,L]$, respectively.  For $M_1$ transmitting and $M_2$ receiving antennas, {the $M_2\times M_1$ time-domain baseband channel impulse response matrix at time $t'$ can be represented as}
\begin{align}
\tilde{\Ht}(t')=\sum_{\ell=1}^L b_{\ell} \delta(t'-\tau_{\ell})e^{j2\pi f_{D,\ell} t'} \av(M_2,\phi_{\ell})\av^T(M_1,\theta_{\ell}),
\label{eq-Ht1}
\end{align}
where for the $\ell$-th multipath, {the sensing parameters $b_{\ell}$ is its amplitude of complex value accounting for both signal attenuation and initial phase difference},  $\tau_{\ell}$ is the propagation delay, and $f_{D,\ell}$ is the associated Doppler frequency, and $\otimes$ denotes the Kronecker product. {Strictly speaking, the amplitude $b_{\ell}$ is frequency dependent. For typical cellular systems where the fractional bandwidth (signal bandwidth normalized to carrier frequency) is small, the variation of $b_{\ell}$ across the whole bandwidth is small and hence we assume it is frequency independent here.} For sensing, $\{\tau_{\ell},f_{D,\ell},\phi_{\ell}, \theta_{\ell}, b_{\ell}\}$ are the \textit{sensing parameters} to be estimated from (\ref{eq-Ht1}). We define a \textit{channel static period} when all these parameters maintain almost unchanged, which is typically a few milliseconds (equivalent to the length of hundreds of OFDM symbols). 

{Equation (\ref{eq-Ht1}) represents the channel impulse response that can be used for both communication and sensing. Note that for communications, we generally only need to know the composited values of the matrix $\tilde{\Ht}$, which are typically obtained by directly estimating some elements in the channel matrix and obtaining the rest via interpolation. For radio sensing, however, the system needs to resolve the detailed channel structure and estimate the sensing parameters. For extended sensing primarily based on machine learning techniques \cite{8067693}, these parameters may be not explicitly needed, which is beyond the scope of this paper.}

The received signal is converted to frequency domain for processing. For the $t$-th OFDM block, the frequency-domain channel matrix at the $n$-th subcarrier corresponding to  (\ref{eq-Ht1}) is given by
\begin{align}
\Ht_{n}=&\sum_{\ell=1}^L b_\ell e^{-j2\pi n\tau_{\ell}f_0}e^{j2\pi t f_{D,\ell}T_s}\av(M_2,\phi_{\ell})\av^T(M_1,\theta_{\ell}),
\label{eq-hfbasic}
\end{align}
where we have approximated the Doppler phase changes over the samples in one OFDM block as a single value. We will work on several slightly varied versions of (\ref{eq-hfbasic}), but still denote them as $\Ht_n$ to show their connections.

\subsection{Formulation for Downlink Sensing}
As mentioned in Section \ref{sec-updown}, sensing is done for the channel environment between each transmitter and receiver, through the echoes specific to the environment. Hence for both downlink and uplink sensing, we separately formulate and use the signals received at each RRU. It is possible to jointly process the signals for sensing, but the benefits are not obvious unless the channels are highly correlated. This could be a significant difference between communication and sensing.

For downlink sensing, each RRU sees reflected downlink signals from itself and the other $Q-1$ RRUs. Its received signal at the $n$-th subcarrier and the $t$-th OFDM block can be represented as
\begin{align}
\yf_{n,t}=&\sum_{q=1}^Q\sum_{\ell=1}^{L_q} b_{q,\ell} e^{-j2\pi n \tau_{q,\ell}f_0}e^{j2\pi t f_{D,q,\ell} T_s}\cdot\notag\\
\label{eq-yf1}
& \quad\av(M,\phi_{q,\ell})\av^T(M,\theta_{q,\ell})\xf_{q,n,t}  +\mathbf{z}_{n,t},\\
=&\underbrace{\A(M,\bm{\phi})\C_n\D_{t}\U^T}_{\Hf_n}\xf_{n,t}+\mathbf{z}_{n,t},
\label{eq-active}
\end{align}
where variables with subscript $q$ are for the $q$-th RRU, $\xf_{q,n,t}$ are the transmitted signals at subcarrier $n$ from the $q$-th RRU, 
\begin{align}
&{\A(M,\bm{\phi})}={(\A_1(M,\bm{\phi}_1),\cdots,\A_Q(M,\bm{\phi}_Q))},\\
&{\xf_{n,t}}={(\xf_{1,n,t},\cdots,\xf_{Q,n,t})^T},\\
&\U=\text{diag}\{\A_1(M,\bm{\theta}_1), \A_2(M,\bm{\theta}_2),\cdots,\A_Q(M,\bm{\theta}_Q)\}, 
\label{eq-U2}
\end{align}
and hence $\U$ is a $MQ\times L$ block diagonal matrix. The $\ell$-th column in $\A_q(M,\bm{\phi}_q)$ (or $\A_q(M,\bm{\theta}_q)$) is $\av(M, \phi_{q,\ell})$ (or $\av(M,\theta_{q,\ell})$), $\D_t$ and $\C_n$ are diagonal matrices with the $\ell$-th diagonal element being $b_\ell e^{j2\pi tf_{D,\ell} T_s}$ and $ e^{-j2\pi n \tau_{\ell}f_0}$, respectively, $\mathbf{z}_{n,t}$ is the noise vector. The model in (\ref{eq-yf1}) has a similar channel structure representation with the basic one in (\ref{eq-Ht1}), but specifies multipath signals to different RRUs. 

According to (\ref{eq-active}), we can see that packing $\yf_{n,t}$ from multiple RRUs can increase its length, but the unknown parameters are similarly increased. Hence sensing does not directly benefit from jointly processing. However, due to channel reciprocity, parameters for signal propagation between RRUs could be similar. Such a property can be exploited for joint processing across RRUs.

\subsection{Formulation for Uplink Sensing}\label{sec-direct}

The received signal in a RRU at the $n$-th subcarrier and the $t$-th OFDM block can be represented as

\begin{align}
\yf_{n,t}&=\sum_{k=1}^K\sum_{\ell=1}^{L_k} b_{k,\ell} e^{-j2\pi n \tau_{k,\ell}f_0}e^{j2\pi t f_{D,k,\ell} T_s}\cdot\notag\\
& \quad\av(M,\phi_{k,\ell})\av^T(M_T,\theta_{k,\ell}) \xf_{k,n,t}  +\mathbf{z}_{n,t},
\label{eq-yfnt0}
\end{align}

Comparing (\ref{eq-yfnt0}) with (\ref{eq-yf1}), we can see that they have similar expressions except for different symbols and parameter values. {Hence, next we will develop a common on-grid expression for both downlink and uplink sensing. }

\subsection{Generalized Delay-Quantized On-grid Formulation}\label{sec-ongrid}

 Let $N_u$ and $\mathsf{S}$ be the number and index set of available subcarriers for sensing, respectively. For downlink sensing, $N_u=N$. We assume that $N\gg L$ and $N$ is large enough such that the quantization error of $\tau_{\ell}$ is small and the delay estimation can be well approximated as an on-grid estimation problem. Let the delay term $e^{-j2\pi n\tau_\ell f_0}$ be quantized to $e^{-j2\pi n\ell/(gN)}$, where $g$ is a small integer and its value depends on the method used for estimating $\tau_{\ell}$. The minimal resolvable delay is then $1/(gB)$. 

Let $K$, $M$ and $M_T$ denote the total number of users/RRUs, the number of antennas for sensing, and the number of antennas in each user/RRU for transmitting, respectively, for either uplink or downlink sensing. We now convert the multipath signal models in (\ref{eq-yfnt0}) and (\ref{eq-yf1}) to a generalized on-grid (delay only) sparse model, by representing it using $N_p\gg L, N_p\le gN$ multipath signals where only $L$ signals are non-zeros. Referring to (\ref{eq-active}) and (\ref{eq-yfnt0}), {the generic delay-on-grid model, applicable to further processing for either downlink sensing or uplink sensing can be represented as, }
\begin{align}
\yf_{n,t}=\underbrace{\A(M,\bm{\phi})\C_n\D_{t}\Pc\U^T}_{\Hf_n}\xf_{n,t}+\mathbf{z}_{n,t}.
\label{eq-yf4}
\end{align}
{Note that to show the connection, we used similar symbols here with some of those in (\ref{eq-active}), however, the definitions are slightly different.} Here, $\C_n$ is now redefined as $\C_n=\diag\{e^{-j2\pi n/(gN)},\cdots, e^{-j2\pi n N_p/(gN)}\}$, re-ordered according to the quantized delay values; $\Pc$ is a $N_p\times L$ rectangular permutation matrix that maps the signals from a user/RRU to its multipath signal, and has only one non-zero element of value $1$ in each row; the other symbols have similar expressions with those in (\ref{eq-active}), with elements in $\A(M,\bm{\phi})$ and $\D_{t}$ being reordered according to the delay. More specifically, the columns in $\A(M,\bm{\phi})$ of size $M\times N_p$ and the diagonal elements in $\D_{t}$ of size $N_p\times N_p$ are now re-ordered and tied to the multipath delay values. $\U$ is an $M_TK\times L$ block diagonal radiation pattern matrix for $M_T$ arrays. $\xf_{n,t}$ is the $M_TK\times 1$ symbol vector. For the moment, we allow repeated delay values in $\C_n$ to account for multipath signals with the same quantized delay but different AoAs and/or AoDs.

\subsection{Selection of Compressive Sensing Algorithms}\label{sec-grid selection}
	
Recently, compressive sensing (CS) techniques have been widely applied in radar sensing \cite{Hadi2015} as well as in JCAS systems \cite{Zhang17basic}. The five sensing parameters in (\ref{eq-hfbasic}) can be estimated either individually or jointly by forming from 1D to 4D CS models. In this paper, we propose to use 1D compressive sensing based on the on-grid formulation in (\ref{eq-yf4}), as will be detailed in Section \ref{sec-onlysymbol} and \ref{sec-channelsymbol}, mainly for the following three reasons:
{
\begin{itemize}
	\item Although high-dimensional on-grid CS algorithms such as the Tensor tool and Kronecker CS \cite{Caiafa13} could offer better performance when there are sufficient measurements in each dimension, they could face large quantization errors in the domains of Doppler frequency, AoD and AoA for our problem here, due to the limited number of measurements associated with short channel coherent time and small amount of antennas. Comparatively, the cellular signals generally have hundreds to thousands of subcarriers, which provide numerous measurements for the delay. Therefore, quantizing delay only can potentially lead to smaller errors. 
	\item Off-grid CS algorithms are yet to be extended to high-dimensional problems, and MMV and block-CS models. There exist some CS techniques dealing with off-grid models, such as the perturbation approach \cite{0266-5611-29-5-054008} and atomic norms \cite{6576276}. But they have high complexity and also have respective constrains on the parameter estimation range and the minimum separation of the parameter values. 
	\item Our 1D methodology provides a solid basis for future extension.  Generally, higher-dimensional CS algorithms can achieve better estimation performance, but they also involve much higher computational complexity. Our 1D methodology provides a path for many potential extensions, for example, replacing the 1D on-grid model with a 1D off-grid model, should off-grid algorithms be extended to the MMV models. 
\end{itemize}
}

\section{Direct Estimation of Sensing Parameters}
\label{sec-onlysymbol}

We now propose a scheme based on 1D CS for estimating the spatial parameters directly using the signal $\yf_{n,t}$ in (\ref{eq-yf4}). This scheme works for all the three sensing methods. We assume that the symbols $\xf_{n,t}$ are known and $N\gg L$. For uplink sensing, this can be achieved by demodulating the symbols as sensing can tolerate more delay than communication, while for downlink sensing, they are centrally known. Note that, the range and indexes of subcarriers in downlink and uplink sensing could be different. RRUs can see signals at more subcarriers in downlink sensing than uplink because in the uplink the total subcarriers could be shared by different group of users and channels are specific to each user.

{We first organize the received signals to a form such that from it 1D CS algorithms can be applied to get the estimates for the delay. From the associated amplitude estimates corresponding to the delay estimates, we then retrieve other sensing parameters.}

Rewrite (\ref{eq-yf4}) as
\begin{align}
\yf_{n,t}^T 
&=\xf^T_{n,t}(\cv^T_n\otimes\I_{M_TK}) \V\A^T(M,\bm{\phi}).
\end{align} 
where $\cv_n=(e^{-j2\pi n/(gN)},\cdots, e^{-j2\pi N_p/(gN)})^T$, $\I_{M_TK}$ is an $M_TK\times M_TK$ identity matrix, and $\V$ is a $M_TKN_p\times N_p$ block diagonal matrix
\begin{align}
\V=\text{diag}\{b_\ell e^{-j2\pi t f_{D,\ell}T_s}\U\p_\ell\}_{\ell=1,\cdots,N_p},
\label{eq-blockv}
\end{align}
with $\p_\ell$ being the $\ell$-th column of $\Pc^T$. 

We have now separated signals $\xf^T_{n,t}(\cv^T_n\otimes\I_{M_TK})$ that are known and dependent on $n$ from other parameters. Then we can stack all row vectors $\yf_{n,t}^T, n\in \mathsf{S}$ to a matrix\footnote{In uplink sensing, there may be less than $N$ vectors available and they may be dis-continuous in index.}, and obtain

\begin{align}
\Yf_t&\triangleq(\yf_{1,t},\cdots,\yf_{n,t},\cdots)^T 
= \W\V\A^T(M,\bm{\phi}),
\label{eq-vat}
\end{align}
where $\W$ is a $N_u\times M_TKN_p$ matrix with its $n$-th row being $\xf^T_{n,t}(\cv^T_n\otimes\I_{M_TK})$.

Inspecting (\ref{eq-vat}), we can see that the estimation problem in (\ref{eq-vat}) can be treated as a multi-measurement vector (MMV) block sparse problem \cite{LIU201480} with $N_u\times M$ observations $\Yf_t$, sensing matrix $\W$, and block sparse signals $\V\A^T(M,\bm{\phi})$ of $L$-sparsity. Let $\V=(\V_1^T,\V_2^T,\cdots,\V_{N_p}^T)^T$ where $\V_\ell$ denotes the $M_TK\times N_p$ block signals, and $L$ out of $N_p$ $\V_\ell$s have non-zero elements. {The non-zero rows and their values in $\V\A^T(M,\bm{\phi})$ can then be solved by various MMV CS algorithms, such as the fast marginalized block sparse Bayesian learning algorithm (BSBL-FM) in \cite{LIU201480,Wipf07} that is adopted in this paper. The indexes of the non-zero rows correspond to the quantized delay values. Their amplitude values can be further used to estimate other sensing parameters. }

{The detailed estimation process based on $\V\A^T(M,\bm{\phi})$ is described for two cases next. We first consider a simple case when there is only one multipath at each delay value. In this case, a simple estimation algorithm is available for estimating all sensing parameters. We then extend the solution to the case when there are multiple multipath signals at each quantized delay bin. We will show that when these multipath signals are from different RRUs, the parameters can be similarly estimated to those in the single multipath case. Otherwise, more complex techniques need to be applied. The method for separating the two cases is yet to be developed. }

\subsection{Single Multipath for Each Delay}\label{sec-singlem}

We first consider the noiseless case. Once the $L$ nonzero blocks $\V_\ell\A^T(M,\bm{\phi})$ are {obtained by BSBL-FM}, we can then get the $L$ delay estimates according to the indexes of the blocks.
 
From (\ref{eq-blockv}) we can see that only the $\ell$-th column in $\V_\ell$ has non-zero elements $b_\ell e^{-j2\pi t f_{D,\ell}T_s}\U\p_\ell$ if $b_\ell\neq 0$. Therefore, 
\begin{align}
\V_\ell\A^T(M,\bm{\phi})=b_\ell e^{-j2\pi t f_{D,\ell}T_s}\U\p_\ell\av^T(M,\phi_\ell).
\end{align}
Since $\p_\ell$ only has a single non-zero element $1$, $\U\p_\ell$ will generate a column vector corresponding to one column in $\U$. Because $\U$ is a block diagonal matrix, only $1$ out of $K$ $M_T\times 1$ vectors in each column is non-zero. 

Now represent $\V_\ell\A^T(M,\bm{\phi})$ as $K$ $M_T\times M$ sub-matrices $(\B_{\ell,1}^T,\cdots,\B_{\ell,K}^T)^T$. If $\B_{\ell,k}\neq 0$, then this multipath is from the $k$-th RRU (user).  We can also see that 
\begin{align}
\B_{\ell,k}=b_\ell e^{-j2\pi t f_{D,\ell}T_s}\av(M_T,\theta_{k,\ell})\av^T(M,\phi_{k,\ell}).
\label{eq-vat23}
\end{align}
From $\B_{\ell,k}$, calculating the cross-correlation between columns and rows, we can obtain AoA or AoD estimates, depending on the order of calculation. { Let $(\B_{\ell,k})_{\cdot,p}$ and $(\B_{\ell,k})_{\cdot,q}$ denote the $p$-th column and $q$-th row of $(\B_{\ell,k})$, respectively. We then have
	\begin{align}
	&\sin(\phi_{k,\ell})\approx\frac{1}{\pi}\angle\left(\sum_{p=1}^{M-1}((\B_{\ell,k})_{\cdot,p})^*(\B_{\ell,k})_{\cdot,p+1}\right),\notag\\
	&\sin(\theta_{k,\ell})\approx\frac{1}{\pi}\angle\left(\sum_{q=1}^{M_T-1}((\B_{\ell,k})_{q,\cdot})^*(\B_{\ell,k})_{q+1,\cdot}\right).
	\end{align}
}

The Doppler frequency $f_{D,\ell}$ can be estimated across multiple OFDM blocks, based on the cross-correlation of $\B_{\ell,k}$ in these blocks: { Let $\B_{\ell,k,t}$ denote the $\B_{\ell,k}$ obtained from the $t$-th OFDM block signal $\Yf_t$, and $T_d$ be the total OFDM blocks used for estimating the Doppler frequency, then we get
\begin{align}
f_{D,\ell}\approx\frac{1}{2\pi T_s}\angle\left(\sum_{t=1}^{N_d-1}(\B_{\ell,k,t})(\B_{\ell,k,t+1})^*\right).
\end{align}
}
The absolute value of $b_\ell$ can be estimated as the mean power of all elements in $\B_{\ell,k}$. A better estimate is to use the cross-correlation output for estimating AoA. {That is
\begin{align}
|b_{\ell}|^2\approx \left|\sum_{p=1}^{M-1}((\B_{\ell,k})_{\cdot,p})^*(\B_{\ell,k})_{\cdot,p+1}\right|.
\end{align}
}		

In noisy cases, we can sort the blocks $\V_\ell\A^T(M,\bm{\phi}), \ell=1,\cdots,N_p$ according to the estimates of $|b_\ell|$ and use a threshold to filter out blocks corresponding to multipath signals. This threshold can be set with reference to the expected received energy for that delay value, using the path loss model. We can also keep the estimated results for a subset of $N_p$ with larger estimated $b_\ell$s, and then apply data fusion techniques over all measurements over a segment of space, time and frequency domains to get synthesized sensing results.

\subsection{Multiple Multipath Signals with the Same Delay}\label{sec-multiplepath}
We consider the case where there are two multipath signals with the same delay. The analysis below can be easily extended to more general scenarios. Let $\cv_n=(\cv_{n,1}^T,\cv_{n,2}^T,\cv_{n,2}^T)^T$, where $\cv_{n,2}$ represents the repeated entries. We can accordingly represent $\W=(\W_1,\W_2,\W_2)$ and $\V=(\V_1^T,\V_2^T,\V_3^T)^T$ in (\ref{eq-vat}). Then we have
\begin{align}
\W\V\A^T(M,\bm{\phi})=(\W_1,\W_2)\left(\begin{array}{cc}
\V_1\A^T(M,\bm{\phi})\\
(\V_2+\V_3)\A^T(M,\bm{\phi})
\end{array}\right).
\end{align}  
This shows that we can always use a $\cv_n$ with single entry for each quantized delay, and multiple signals with different angles will show up in the MMV estimates. More specifically, if $\ell\in \mathcal{S}$ multipath signals have the same delays but different AoAs or AoDs, we will then get
\begin{align}
\V_\ell\A^T(M,\bm{\phi})=\sum_{\ell\in \mathcal{S}}b_\ell e^{-j2\pi t f_{D,\ell}T_s}\U\p_\ell\av^T(M,\phi_\ell).
\end{align}

If these multipath signals are from different RRUs (users), multiple $\B_{\ell,k}$s will be non-zero. Hence in this case, sensing parameters for these multipath signals can be estimated using the algorithms similar to those for the single multipath case in section \ref{sec-singlem}.

If multipaths are from the same RRU (user), we will have
\begin{align}
\hat{\B}_{\ell,k}=\sum_{\ell\in \mathcal{S}} b_\ell e^{-j2\pi t f_{D,\ell}T_s}\av(M_T,\theta_{k,\ell})\av^T(M,\phi_{k,\ell}).
\label{eq-Bq2}
\end{align}
Obtaining solution from (\ref{eq-Bq2}) is a complicated estimation problem. When the number of multipath with the specific delay value is small, which is a typical scenario, the AoAs and AoDs can be estimated by applying 2D spectrum analysis techniques to each $\hat{\B}_{\ell,k}$, such as by 2D-ESPRIT or 2D-MUSIC algorithm. Across multiple OFDM blocks, 3D spectrum analysis techniques could be applied to additionally get the estimate for Doppler shift too. 
 
\section{Indirect Estimation Using Signal Stripping}
\label{sec-channelsymbol}
 We have seen in the previous section that due to the multiuser-MIMO signal, block CS is applied to estimate the sensing parameters. It has high complexity, and is sensitive to system imperfections, such as quantization errors in the delay. In this section, we propose another  sensing parameter estimating scheme called \textit{signal stripping},  which derives simpler signal models from the received signals, and then estimates parameters based on these simplified models. 
 
\subsection{Signal Stripping}

The idea of signal stripping is to simplify the signal input to sensing algorithms by removing the modulated symbols from the signal and separating channels for different nodes (MSs for uplink sensing or RRUs for donwlink sensing).  More specifically, this approach uses the estimated data symbols and channels to strip signals from different nodes, and keep as few as a single node’s composited channel matrix (with estimated elements in the channel matrix) as input to sensing. This method can significantly reduce the number of sensing parameters to be estimated each time, reduce the algorithm complexity and improve its performance, should the estimated composited channel matrix for each node be accurate. 

Referring to (\ref{eq-active}) or (\ref{eq-yf1}), the key is to get an accurate frequency-domain channel matrix estimate at subcarrier $n$, at time $t$ for user $k$ 
\begin{align}
\hat{\Ht}_{n,k,t}=\Ht_{n,k,t}+\bm{\Delta}_{n,k,t},
\label{eq-hatht}
\end{align} 
where $\bm{\Delta}_{n,k,t}$ is the channel estimation error, and
\begin{align}
\Ht_{n,k,t}=\sum_{\ell=1}^{L_k} b_{k,t,\ell} &e^{-j2\pi n \tau_{k,t,\ell}f_0}e^{j2\pi t f_{D, k,t,\ell} T_s}\cdot\notag\\
& \av(M,\phi_{k,t,\ell})\av^T(M_T,\theta_{k,t,\ell}).
\label{eq-Ht}
\end{align}
 
In this paper, we do not provide detailed algorithms for refining the composited channel estimation, but present a general approach. The impact of channel estimation error on the performance of sensing will be evaluated in Section \ref{sec-simuind}.

The composited channel matrix in (\ref{eq-hatht}) can be efficiently obtained by estimating and refining the composited channels. Channel matrices in communications are generally estimated with the assistance of interpolation techniques and hence its accuracy is insufficient for estimating sensing parameters. Since sensing can tolerate much larger processing delay than communication, we can exploit the demodulated signals in communication to reconstruct composited channel matrix. This can be implemented in a similar process to the decision directed channel estimation (DDCE) scheme in communication systems \cite{RN28}. Different to conventional DDCE algorithms applied for communications, we only need to reconstruct channels as accurate as possible, but do not need to do channel prediction.

We can reconstruct data symbols after decoding the whole packet in communications, and then use them to get multiple channel estimates during this period. This is because sensing can tolerate delay up to a few milliseconds, which can include tens of packets. 

\subsection{ Estimation of Sensing Parameters}

When the channel estimates in (\ref{eq-hatht}) are obtained, we can use them as inputs to sensing algorithms, and get the estimates for sensing parameters for each user.  This is a typical mathematical model in radar signal processing \cite{6850183}. Here we only consider  the case when there is only one multipath signal within each quantized delay bin for each user, and propose a 1D CS based algorithm for sensing parameter
estimation. The algorithm here can also be applied to the case when only the received signals at pilots such as the DMRS in 5G NR are used for sensing, since these pilots are typically orthogonal for different MSs.

We use (\ref{eq-Ht}) as a generalized channel matrix model, and drop the subscripts $t$ and $k$ in the parameter variables. Referring to Section \ref{sec-ongrid} we consider a similar delay-on-grid model where the delays $\tau_{\ell}f_0$ are quantized as $q_{\ell}/N'$ with $q_{\ell}$ being an integer and $N'=gN$. Therefore $e^{-j2\pi n\tau_\ell f_0}\approx e^{-j2\pi nq_{\ell}/N'}$. This delay-on-grid model for reconstructed channel matrix can be written as
\begin{align}
\Ht_{n}=\A_R\D\C_n\A^T_T,
\label{eq-htnq}
\end{align}
where the $\ell$-th column in $\A_R$ (or $\A_T$) is $\av(M, \phi_{\ell})$ (or $\av(M_T,\theta_{\ell})$), $\D$ and $\C_n$ are diagonal matrices with the $\ell$-th diagonal element being $b_\ell e^{j2\pi tf_{D,\ell} T_s}$ and $ e^{-j2\pi n q_\ell/N'}$, respectively.
The $m$-th column of $\Ht_n$ can be represented as
\begin{align}
\hct_{n,m}=\A_R\D\Pc_{m}\cv_n,
\end{align}
where $\Pc_{m}$ is a diagonal matrix with diagonal elements being the $m$-th row of $\A_T$, and $\cv_n$ is an $L\times 1$ vector with the $\ell$-th element $e^{-j2\pi nq_\ell/N'}$. 

We may have a few options to process different columns of $\Ht_n$. For the least, we need two columns so that AoD can be estimated. Use $M_t=2$ as a simple example. Transpose $\hct_{n,m}, m=1,2$ and stack them into a row vector
\begin{align}
(\hct^T_{n,1},\ \hct^T_{n,2})=\cv_n^T\D(\Pc_{1}\A_R^T,\ \Pc_2\A_R^T).
\end{align}
Now stacking similarly formulated row vectors for all usable subcarriers $n\in \mathcal{S}$ together, we can obtain
\begin{align}
\widetilde{\Ht}=\W\underbrace{\D(\Pc_{1}\A_R^T,\ \Pc_2\A_R^T)}_{\bm{G}},
 \label{eq-mmv}
\end{align}
where the $\ell$-th column of the $N_u\times L$ matrix $\W$  is $\{e^{-j2\pi nq_\ell/N'}\}, n\in \mathcal{S}$.

We now convert the multipath signal model in (\ref{eq-mmv}) to a generalized delay-on-grid sparse model, by representing it using $N_p\gg L, N_p\le N'$ multipath signals where only $L$ signals are non-zeros.  Without any prior knowledge of the delay, we can use $N_p=N'$; otherwise, the range of delays can be reduced. We can then treat it as a MMV CS problem and use algorithms such as OMP or Bayesian CS to get the estimate for $\W$ and $\bm{G}$. The dictionary is a partial $N_u\times N_p$ DFT matrix $\F$. When $N_p=N'$, its $q$-th column is $\{e^{j2\pi nq/N'}\}, n\in \mathcal{S}$.

Let $g_{\ell,p}$ be the $(\ell,p)$-th element of $\bm{G}$ in (\ref{eq-mmv}). 
Once the delays and $\bm{G}$ are estimated, we can get the estimates for AoA and AoD through
 \begin{align*}
 & 2\pi d\sin(\phi_\ell)/\lambda=\angle\big(\underbrace{\frac{1}{2M}\sum_{k=0}^1\sum_{p=1}^{M-1} g_{\ell,p+kM}^H g_{\ell,p+1+kM}}_{\varepsilon_\ell}\big),\\
  &  2\pi d\sin(\theta_\ell)/\lambda=\angle\big(\underbrace{\frac{1}{M}\sum_{p=1}^M g_{\ell,p}^H g_{\ell,p+M}}_{\xi_\ell}\big),
  \end{align*}
  respectively. The value of $|b_{\ell}|^2$ can also be obtained easily during the process of computing AoA and AoD, being either $|\varepsilon_\ell|^2$ or $|\xi_\ell|^2$. The estimates of $|b_{\ell}|^2$ can be used to find the effective multipath delay bins in noisy channels, by using a threshold determined, e.g., as a fractional scalar of the maximum power of multipath signals.
  
  The computation above can be readily extended to the case when $M_T> 2$.
  
 This process can be repeated over multiple refined channel estimates over the channel static period. The Doppler shift can then be estimated from the cross-correlation between two or more $\bm{G}$s that are sufficiently spaced in time,  when channel still remains stable except for the Doppler phase terms. Let $\bm{G}_t$ denote the estimate of $\bm{G}$ from the $t$-th refined channel estimates. Using two $\bm{G}$s the estimates of Doppler phase can be obtained as
 \begin{align}
 2\pi f_{D,\ell}T_s=\frac{T_s}{T}\angle\big((\bm{G}_{t+T})_{\ell,\cdot}\, ((\bm{G}_t)_{\ell,\cdot}) ^H\big),
 \end{align}
  where $(\Xt)_{\ell,\cdot}$ denotes the $\ell$-th row of the matrix $\Xt$ and $T$ is the interval of two symbols used for estimating Doppler shift.
 
\section{Clutter Reduction}\label{sec-clutter}

As discussed in Section I, in this paper, we treat echoes with near-zero Doppler frequencies as clutter. Relatively, we call other echoes with non-zero Doppler frequencies as \textit{dynamic multipath}. 

We propose a low-complexity and efficient \textit{Background Subtraction} solution for clutter reduction, which is inspired from the background subtraction method in image processing \cite{SOBRAL20144}. The basic idea is to construct an estimate for clutter by averaging over a long period, and then subtract it from the input to the sensing algorithms.  This requires static sensing parameters for clutter, and signals that are unmodulated or modulated with the same data. Hence it is suitable for the indirect sensing scheme, and can also be applied to the direct sensing scheme, but only when the received signals corresponding to the training signals in each frame is used. In the following, we will present the solution by referring to the indirect sensing method. 

The proposed processing will be applied to the channel matrix at each subcarrier for each user. From the refined channel matrix estimates, we pick up estimates at an interval of $T_h$ seconds, and denote them as 
\begin{align}
\cdots, \Ht(i-1), \Ht(i), \Ht(i+1),\cdots,
\end{align}
where the expression of $\Ht(i)$ is similar to (\ref{eq-htnq}), but $\Ht(i_1)$ and $\Ht(i_2)$, $i_1\neq i_2$ may have different sensing parameters.

We define a recursive equation for estimating the clutter matrix $\bar{\Ht}$
\begin{align}
\bar{\Ht}(i)=\alpha\bar{\Ht}(i-1) +(1-\alpha)\Ht(i),
\label{sec-recur}
\end{align} 
where $\alpha$ is the learning rate (forgetting factor) and the initial one $\bar{\Ht}(1)$ can be either 0 or computed as the average of several initial $\Ht(i)$s.

There is a major difference for background subtraction between radio sensing and image processing. In image processing, the image difference corresponds to pixel variation. However, in radio sensing, both Doppler shifts and variation in sensing parameters cause difference in two channel matrices. This makes the choice of $T_h$ critical in radio sensing.

Consider a Doppler frequency $f_D$. Its corresponding phase shift at $iT_h$ is given by $\exp(j2\pi f_DT_h i)$. When this multipath's other parameters remain unchanged, applying the recursive equation (\ref{sec-recur}) to the whole channel is equivalently to the Doppler phase only. Let $\rho(i)$ and $\exp(j2\pi f_DT_h i)$ replace $\bar{\Ht}(i)$ and $\Ht(i)$ in (\ref{sec-recur}), respectively. Starting from $i=1$, after $p$ recursions we can get
\begin{align}
\rho(p)=e^{j2\pi f_DT_h}\frac{(1-\alpha)(1-\alpha^p e^{j2\pi f_DT_h p})}{1-\alpha e^{j2\pi f_DT_h}}.
\label{eq-rhop}
\end{align}
When $f_D=0, \rho(p)=1-\alpha^p$. To make $\rho(p)$ approach to 1 for $f_D=0$, $p=500$ is approximately needed. 

\begin{figure}[t]
\centering
\includegraphics[width=\figsize]{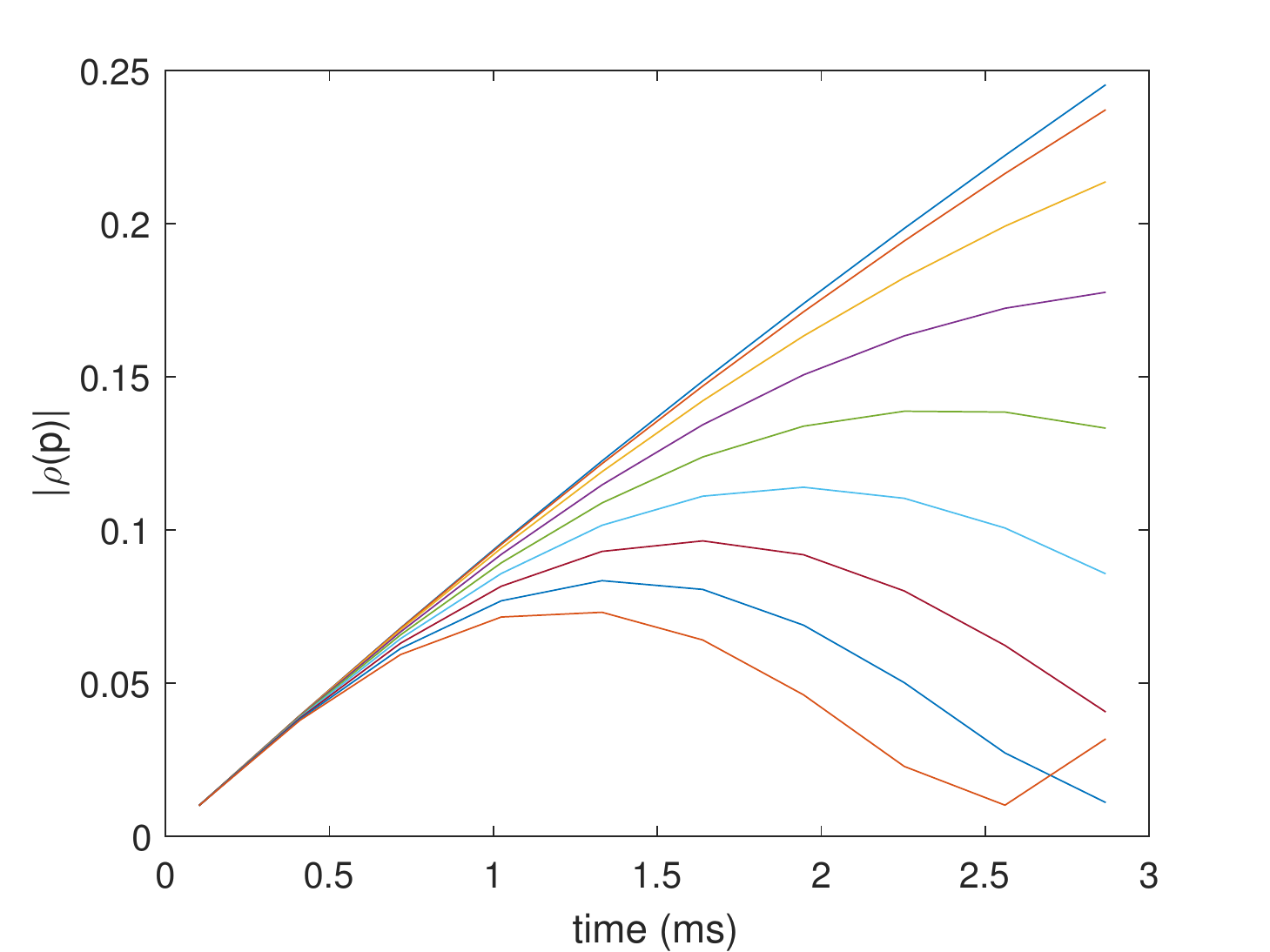}
\caption{ Exemplified values of $\rho(p)$, with $T_h=20T_s$, $\max\{p\}=30$ at approximately 2.8 ms. Curves from top to bottom correspond to Doppler frequencies from $0$ to $400$ Hz at an interval of $50$ Hz. }
\label{fig-rhop}
\end{figure}

As will be detailed in Section \ref{sec-simu}, for typical applications in perceptive mobile networks, the maximum $f_D$ is about $400$ Hz, and the channel stable period is in the order of a few milliseconds. Due to the small Doppler frequency value, the Doppler phases typically change slowly over the channel stable period, unless the vehicle moving speed is very large. This makes averaging at small $T_h$ useless in terms of reducing ``interfering" dynamic multipath signals from the clutter estimation. An example is shown in Fig.\ref{fig-rhop}, which indicates that much larger $T_h$ must be used to get a clear clutter estimate.

\begin{figure}[t]
\centering
\includegraphics[width=\figsize]{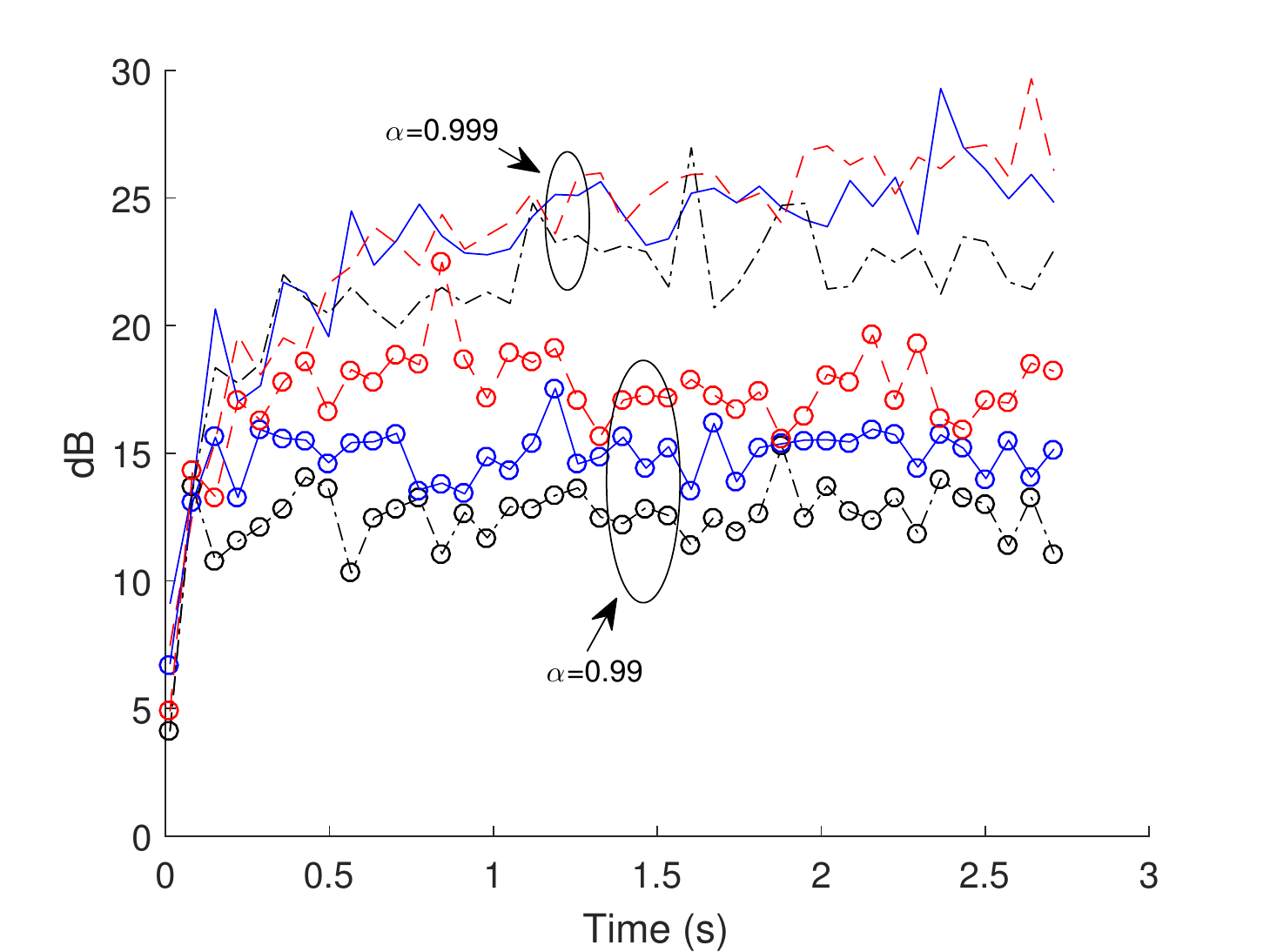}
\caption{ Power ratio of the clutter to ``interfering signals" after applying the recursive averaging, in the presence of approximately 10 non-clutter interfering signals over each period of $270T_s$ (assumed to be channel stable period). Red dashed curves for $T_h=240T_s$, blue solid curves for $T_h=120T_s$, and black curves for $T_h=60T_s$.}
\label{fig-ratio}
\end{figure} 

Since the slowly changing Doppler phases over the channel stable period generally do not cause cancellation for dynamic multipath signals, we want to minimize the number of samples obtained from each channel stable period. However, smaller sampling rate causes slower collection of the clutter signals. Hence a tradeoff is needed here, and the reasonable value is found to be $1$ or $2$ samples per medium channel stable period. A more optimal value may be determined through statistical analysis over a distribution of the Doppler frequency and the channel stable period. On the other hand, the learning rate $\alpha$ also has an important impact on the averaging operation. These effects are demonstrated in Fig. \ref{fig-ratio}. We can see that the power ratio becomes almost stable after 0.5 and 2 seconds of recursive averaging operation for $\alpha=0.99$ and $0.999$ respectively. Larger $\alpha$ achieves better performance.

The clutter estimate is always updated every $T_h$ seconds using (\ref{sec-recur}). Once a stable estimate is obtained, it is subtracted from the current and future refined channel estimates during the interval $T_h$. 

For noisy channel estimates, we can work out the distribution parameters of the combined noise output after the recursion. Assuming the noise in different channel estimates is uncorrelated and each follows the same Gaussian distribution with mean zero and variance $\sigma^2$. Then the output noise matrix will still have zero mean, and covariance matrix $\sigma_c^2\I$, with
\begin{align}
\sigma_c^2&=\sigma^2(1-\alpha^{-1})^2\sum_{i=1}^p (\alpha^2)^i\notag\\
&=\sigma^2(1-\alpha)^2\frac{1-\alpha^{2p}}{1-\alpha^2},
\end{align} 
It can be seen that when $p$ is large, $\sigma_c^2$ approaches to $\sigma^2(1-\alpha)^2/(1-\alpha^2)$. When $\alpha=0.99$, it becomes $0.005\sigma^2$. Hence noise is suppressed in the recursive operation. Our simulation shows that it converges approximately at $p= 150$.

Therefore when subtracting the clutter from the current channel estimate, the noise is almost not increased. This is an advantage of the background subtraction method. In addition, the clutter channel matrix output from the recursive algorithm also allows us to efficiently estimate the clutter parameters. Comparing two estimates obtained at different times, we can also efficiently identify the changes in static objects in the radio image. 

Note that by adjusting the parameters in the recursion equation, we can actually obtain signals with different Doppler frequencies. Hence this method can be extended for separating multipath signals with different Doppler frequencies into different groups. 

\section{Simulation Results}\label{sec-simu}

We present simulation results here to validate the effectiveness of the proposed framework and parameter estimation schemes. { For solving the MMV problems, we use the Block Sparse Bayesian Learning (BSBL) \cite{LIU201480} in direct estimation and Sparse Bayesian Learning (SBL) \cite{4524050} in indirect estimation.} 
 
We consider a system with $4$ RRUs, providing connections to $4$ users through multiuser MIMO. Each RRU has 4 antennas and each MS has 1 antenna. The carrier frequency is $2.35$ GHz and the signal bandwidth is $100$ MHz. Unless stated otherwise, for downlink, all $N=512$ subcarriers are used, and for uplink, 128 subcarriers with random indexes are shared by four users using multiuser-MIMO. No radar cross-section information is considered. 

{The multipath channels are randomly generated in cluster following a complex Gaussian distribution.} We use a pathloss model with pathloss factor 4 for downlink and 2 for uplink sensing. The transmission power of the RRU and MS is 30 dBm and 25 dBm respectively. { Throughout this paper, we assume that the noise is Additive White Gaussian Noise (AWGN) with thermal noise power $N_{0}=-174$ dBm/Hz. Hence the total thermal noise in the receiver is $-174+10\log(10^8)=-94$ dBm.} Multipath signals for each RRU/MS are generated randomly in cluster, mimicking reflected/scattered signals from objects. Multipaths in each cluster are generated following uniform distributions of $[10, 15]$ for the total multipath number, $[0,45]$ degrees for direction span, $[0, 45]$ m for distance, and $[0, 600]$ Hz for Doppler frequency. Across clusters there are additional offsets in direction ($[-75,75]$ degrees), distance ($[50,180]$ m) and moving speed ($[-40,40]$ m/s), reflecting the different locations of the transmitter to the sensing receiver. Unless stated otherwise, delays are on grid with an interval of 10 ns, corresponding to a distance resolution of $3$ m. Delays from the same RRU/MS are kept different. But they could be the same between RRUs/MSs. Random continuous values are used for Doppler shift, AoAs and AoDs. { From the pathloss factors and the multipath propagation distances, we can see that the received signal-to-noise power ratio (SNR) for estimating the sensing parameters for a particular multipath could be as low as $0$dBm for the downlink sensing, while it is much higher ($\geq 30$ dB) for uplink sensing.}

Based on these parameters, we can work out an approximate (minimum) channel stable period that sensing parameters remain unchanged. Assume this period lasts when vehicles/objects move less than 5 cm, and the maximum relative moving speed is 30 m/s. This period is then $0.05/30=0.0017$ s, equivalent to the period of $(0.0017/(512/10^{8}\times1.25)\approx 265$ OFDM blocks.   

In all the figures below, unless stated otherwise, every plus or circle represents parameters for one multipath: Pluses and circles are for estimated and actual ones, respectively. Different colors represent multipath from different RRUs/MSs. In each figure, 10 implementations are plotted. The sensing parameters are fixed in all 10 implementations, but the data symbols and noise are changed. The AoA estimates are shown in the form of AoA phase, $\pi\sin(\phi_{\ell})$, in either degrees or radian.

\subsection{Direct Estimation}\label{sec-simud}

\begin{figure}[t]
\centering
\includegraphics[width=\figsize]{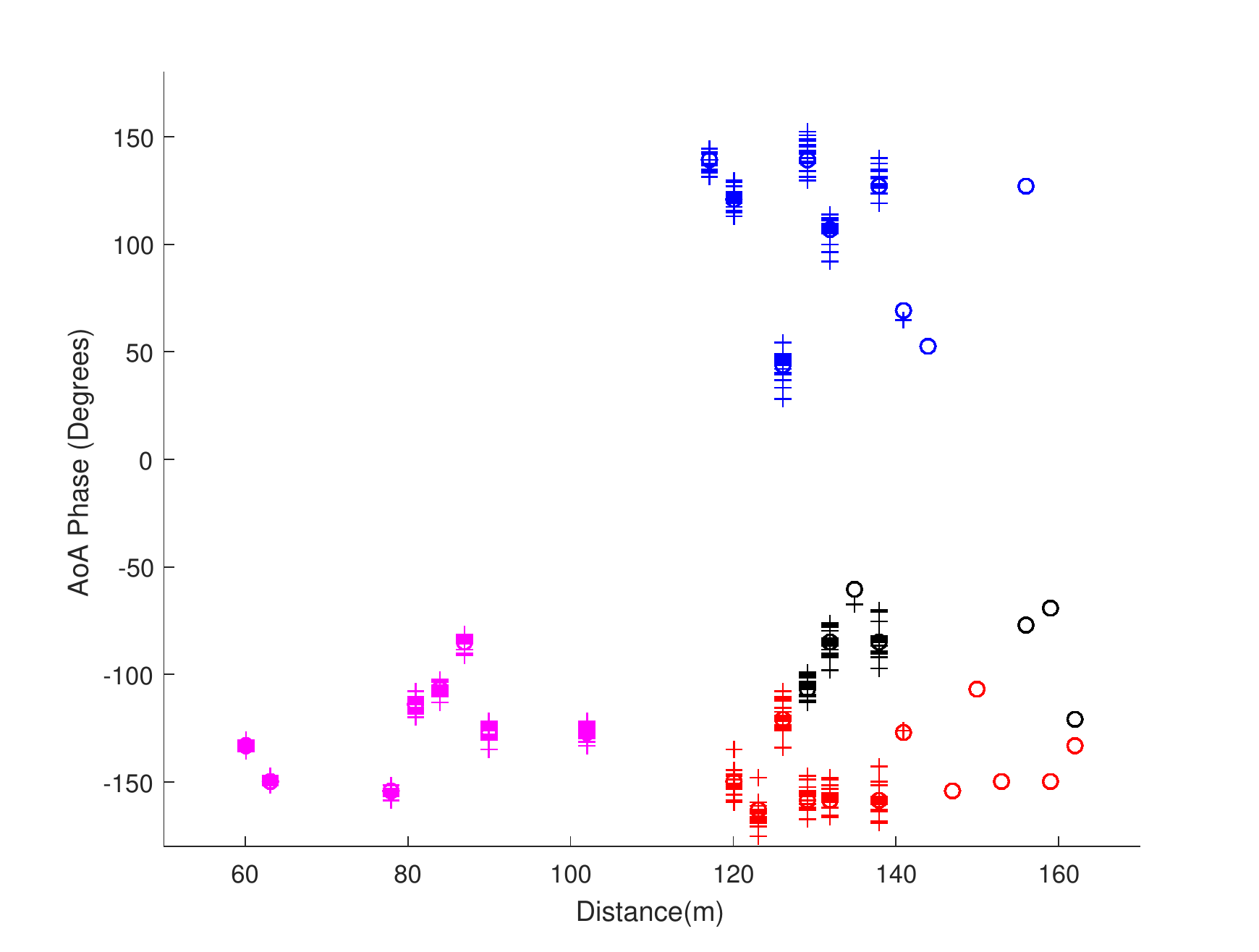}
\caption{10 implementation results for AoA-Distance estimation in direct downlink sensing.  }
\label{fig-down}
\end{figure}

\begin{figure}[t]
\centering
\includegraphics[width=\figsize]{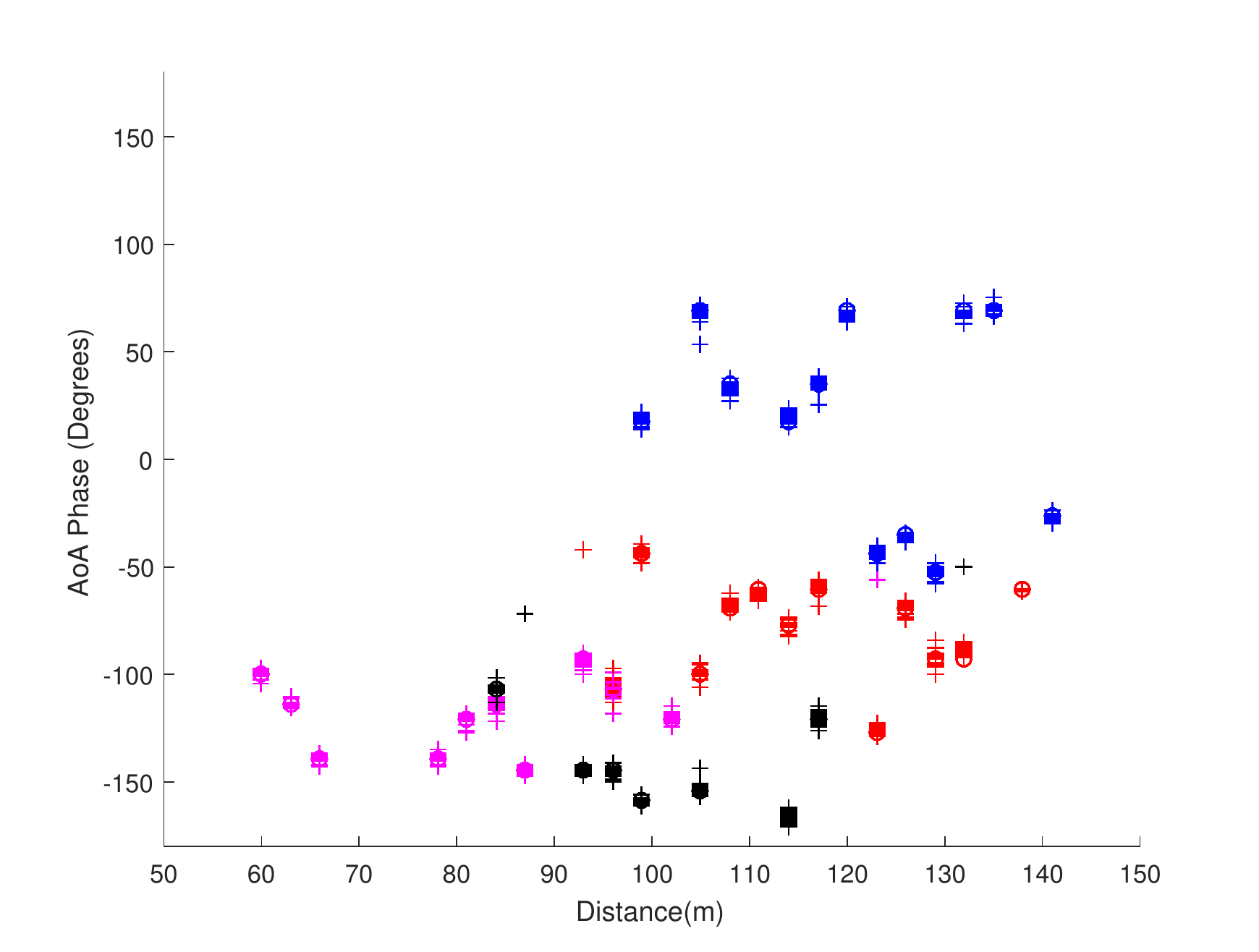}
\caption{10 implementation results for AoA-Distance estimation for direct uplink sensing.  }
\label{fig-up}
\end{figure}

Figs. \ref{fig-down} and \ref{fig-up} present typical AoA-Distance results for downlink and uplink sensing respectively. Note that the depicted distance is the total signal travelling distance between a transmitter and the receiver, and does not necessarily translate to the distance of objects to the receiver directly. Complex across-RRU synthesizing is needed to achieve the translation, particularly for uplink sensing. Both figures demonstrate that the estimates are quite robust and accurate, when the received SNR is sufficiently high. Note that there are no matching estimates for some multipath at distances larger than approximately 145 meters due to the low SNR here. This is particularly obvious in the downlink sensing case where the adopted pathloss factor is 4, compared to 2 in the uplink.

\subsection{Indirect Estimation based on Signal Stripping}\label{sec-simuind}

Instead of implementing the DDCE algorithm to actually refine the channel estimation, we introduce channel reconstruction error in (\ref{eq-hatht}) as AWGN to evaluate the performance for the indirect estimation method. The \textit{signal-to-interference ratio (SIR)} between the mean power of  the channel coefficients and the reconstruction error is denoted by $\eta$.  

\begin{figure}[t]
\centering
\includegraphics[width=\figsize]{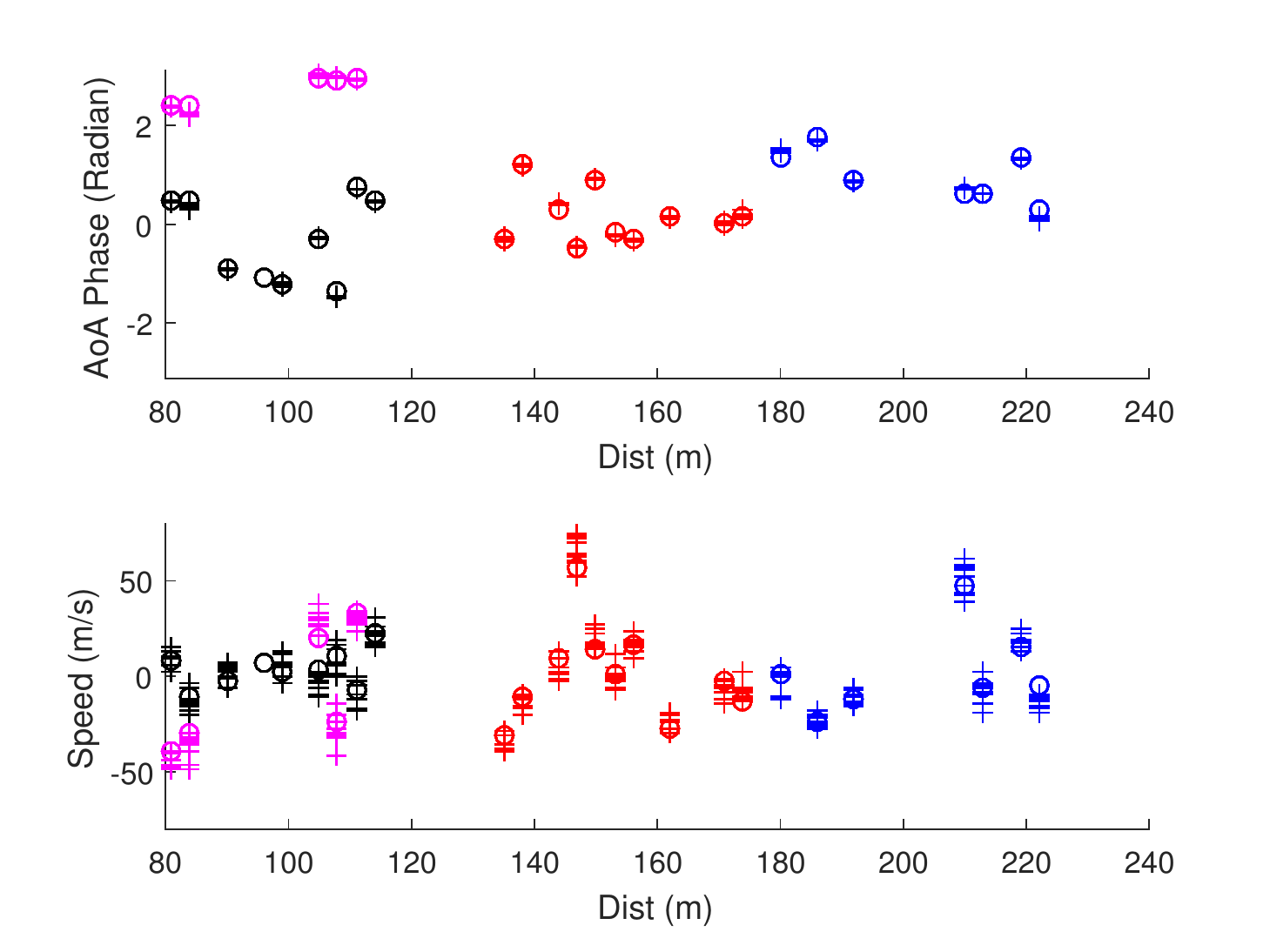}
\caption{10 implementation results for indirect uplink sensing with channel SIR $\eta=15$dB, and symbol sampling interval $T=20T_s$.  }
\label{fig-uplink_ind1}
\end{figure}

Figs. \ref{fig-uplink_ind1} shows the results for uplink sensing. It can be seen that the estimates of delay and AoA are accurate and are robust to the introduced channel reconstruction error. The estimates of moving speed, through estimating the Doppler frequency $ f_{D,\ell}$, have relatively large errors because the actual Doppler phase values are very small and hence sensitive to both noise and the interval $T$.

Comparing the results here and those in \ref{sec-simud}, it is suggesting that the indirect methods can achieve better performance than the direct method when the estimation error in the channel matrix is small enough, as different users' channels/signals are efficiently separated. Hence the key is to develop a low-complexity high-accuracy channel refining algorithm, which is our on-going work.

{
In Fig. \ref{fig-downlink_offgrid1}, we present the results for the case when delay is generated as continuous values (off-grid model) for  downlink sensing. The figures  show that delays and AoAs are identified with degraded but still acceptable accuracy, but the speed estimation varies significantly across different realizations. As a comparison, we also plot the sensing results for AoA-Distance in Fig. \ref{fig-dft_downlink} by applying the classical 2D DFT method to the signal in (\ref{eq-Ht}). For AoA, a 64-point DFT is applied to the 4 signals received at four antennas. The image is cleared by setting 2D-DFT outputs with power 25dB lower than the maximum to zeros. Comparing Fig. \ref{fig-downlink_offgrid1} with Fig. \ref{fig-dft_downlink}, we can see that the proposed 1D CS method achieves much better resolution than the classical 2D DFT method.

\begin{figure}[t]
	\centering
	\includegraphics[width=\figsize]{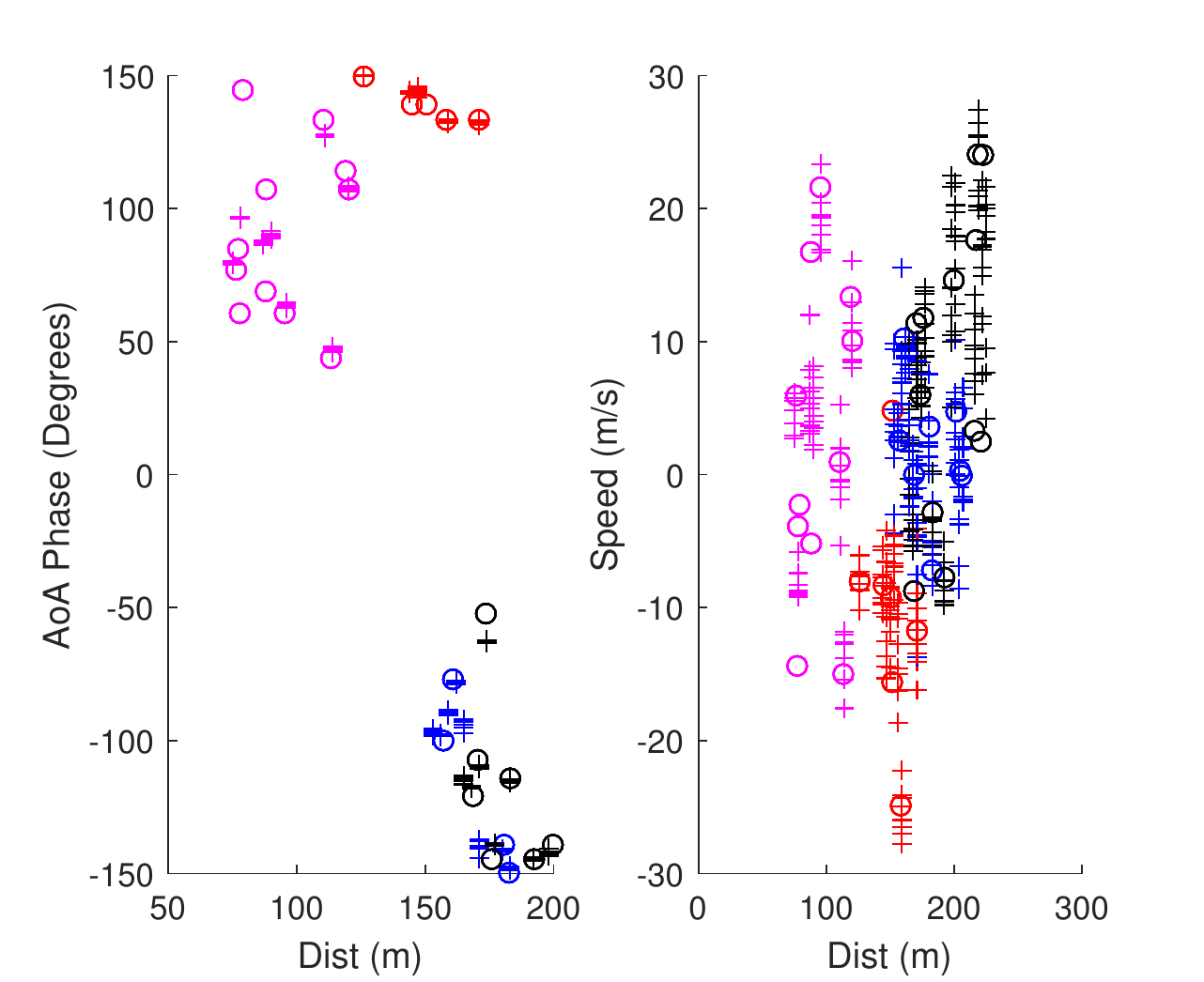}
	\caption{10 implementation results for indirect downlink sensing with $\eta=15$dB, and $T=20T_s$, where delay values are continuous (off-grid). Subcarriers are interleaved. }
	\label{fig-downlink_offgrid1}
\end{figure}

\begin{figure}[t]
	\centering
	\includegraphics[width=\figsize]{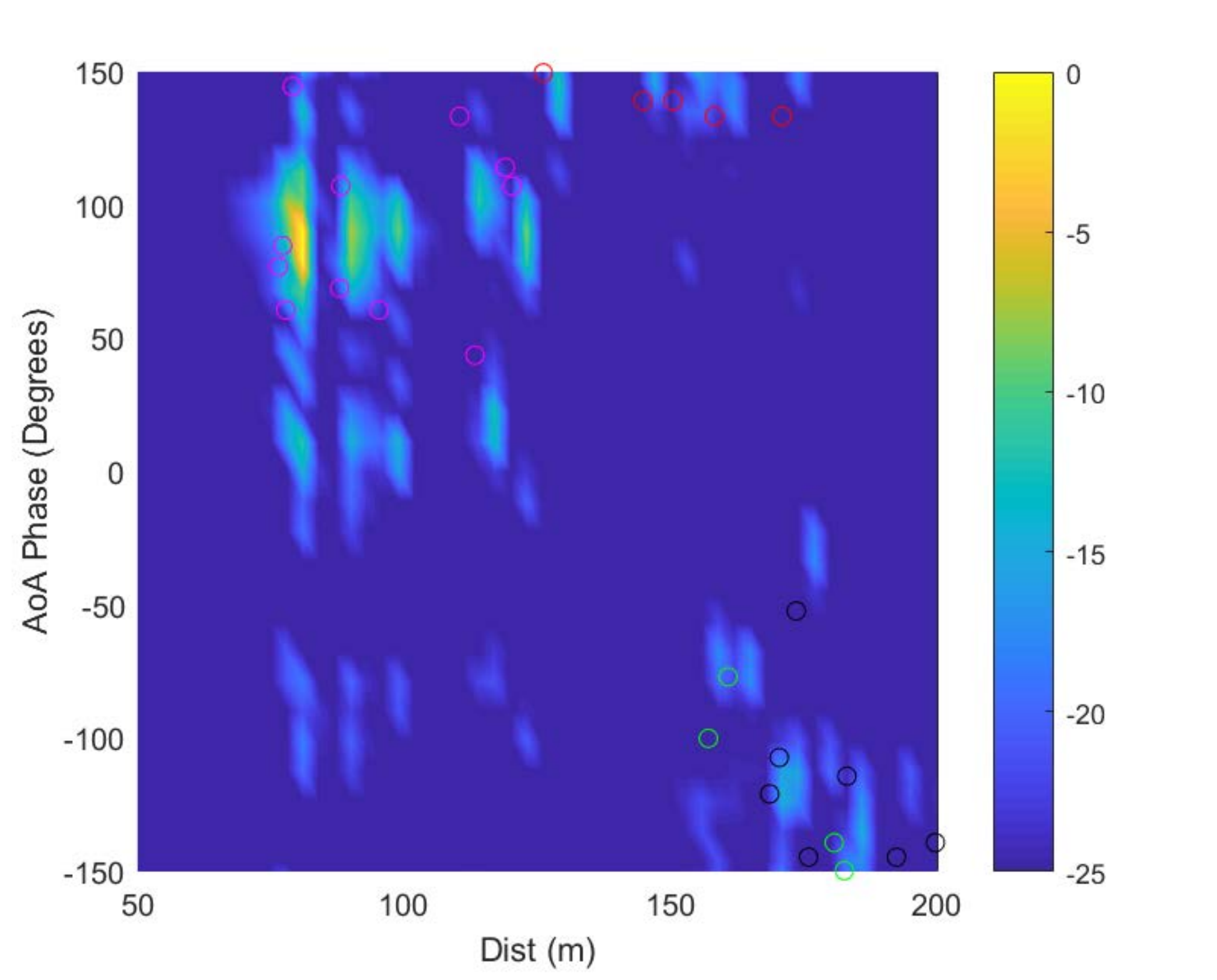}
	\caption{10 implementation results for classical 2D DFT results. System setup is the same as that in Fig. \ref{fig-downlink_offgrid1}.}
	\label{fig-dft_downlink}
\end{figure}
}

\begin{figure}[t]
	\centering
	\includegraphics[width=\figsize]{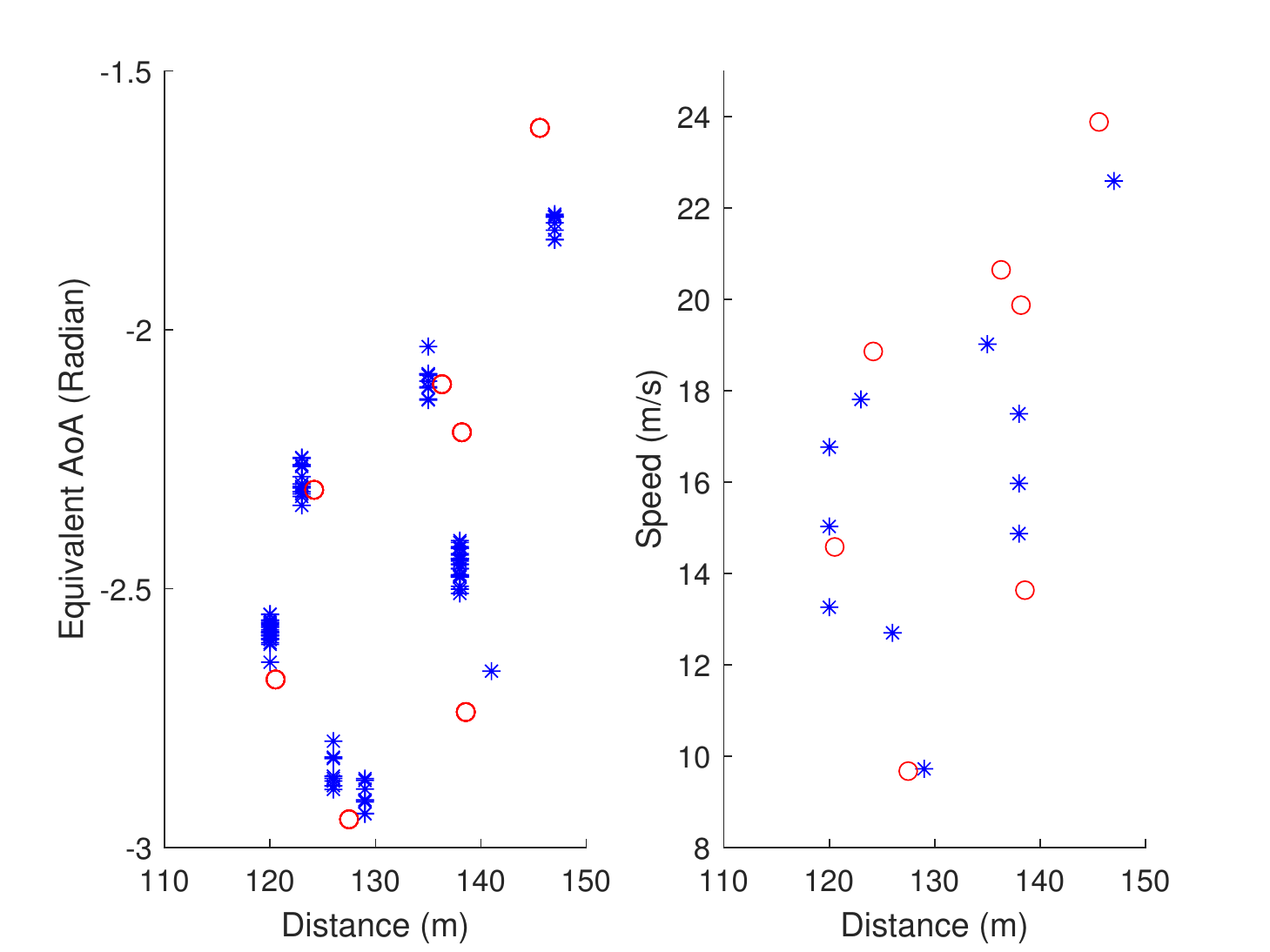}
	\caption{Sensing parameter estimation using the proposed 1D CS method for indirect uplink sensing, with $\eta=15$dB. ``Equivalent AOA'' equals to $\pi\sin(\theta)$, and speed is relative to the static BS. All parameters have continuous values (off-grid model).}
	\label{fig-downlink_dmrs_1d}
\end{figure}

\begin{figure}[t]
	\centering
	\includegraphics[width=\figsize]{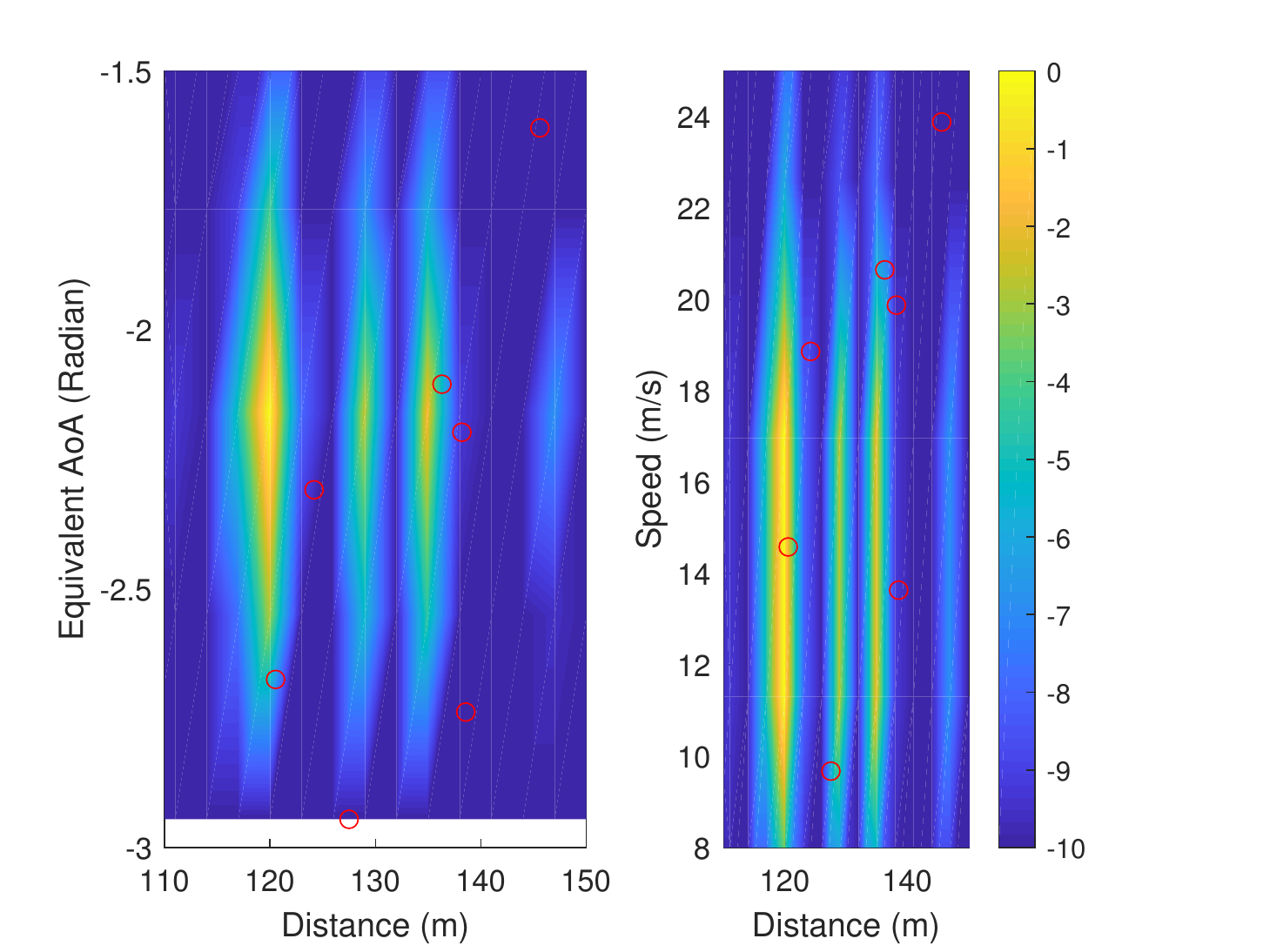}
	\caption{Sensing parameter estimation using the N-way OMP (3D CS) algorithm in \cite{Caiafa13}. Dictionaries for AoA and Doppler estimation are interpolated to size 16 and 256, respectively. Other configurations are similar to those in Fig. \ref{fig-downlink_dmrs_1d}. }
	\label{fig-downlink_tensor}
\end{figure}

{
We further test our indirect uplink sensing scheme using a practical subcarrier allocation example in 5G NR with the Type B set-up and a total subcarriers of 252. Within each resource block of 12 continuous subcarriers, four subcarriers with indexes 3, 4, 9 and 10 are used. Hence a total of 84 subcarriers. The sampling period $T=35T_s$, and a total of 8 sets of observations are obtained for Doppler estimation. For comparison, the N-way OMP Tensor (NOMP) method \cite{Caiafa13}, which is a 3D CS algorithm, is also simulated. The simulation results are presented in Fig. \ref{fig-downlink_dmrs_1d} and Fig. \ref{fig-downlink_tensor} for our scheme and the NOMP scheme, respectively. In both figures, only estimated multipath channels with power within 10dB of the maximum are shown. Comparing these two figures, we can see that the proposed 1D CS method achieves better resolution for distance, AoA and speed for most multipath channels. This demonstrates the effectiveness of our scheme in the cases when  sufficient measurements are only available in the delay domain.
}

\subsection{Effect of Clutter Suppression}

We only present the simulation results for the background subtraction method here, as for the differential method, in the simple form, it is almost a repeat of the simulation in Section \ref{sec-simuind} with increased number of multipath and noise. { The clutter signals are generated similarly to other multipath signals, with Doppler frequencies set to be near-zero values.}

\begin{figure}[t]
\centering
\includegraphics[width=\figsize]{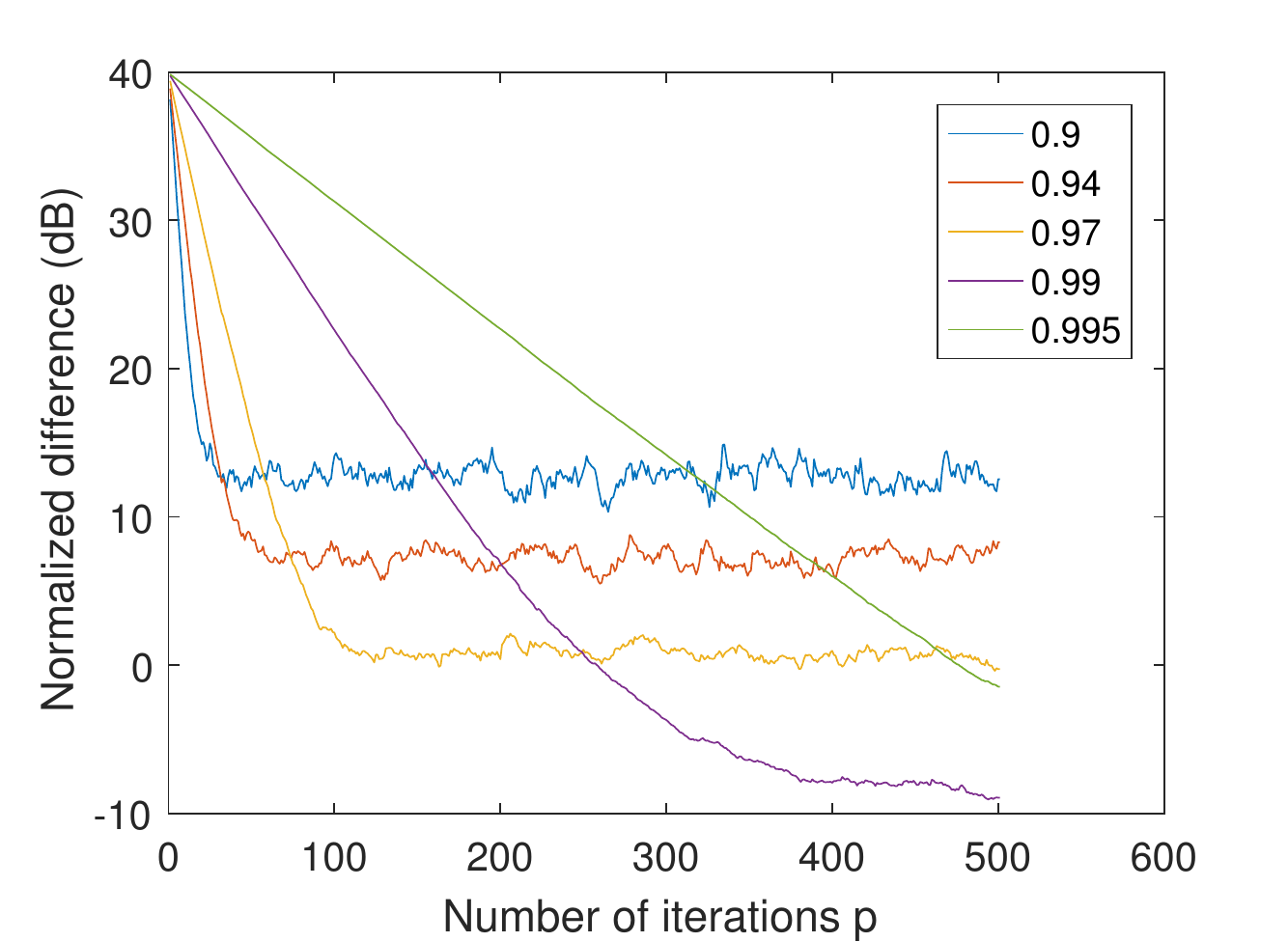}
\caption{Difference between the reconstructed and the true clutter signals, normalized to the power of the true clutter. Learning rates $\alpha$ for curves from left to right are 0.9,0.94,0.97,0.99 and 0.995, respectively. }
\label{fig-cl_diff}
\end{figure}

In Fig. \ref{fig-cl_diff}, we plot the normalized difference between the output from the recursive reconstruction algorithm and the actual clutter. We use one sampled channel estimate within each channel stable period of 2 ms. The figure indicates that $\alpha>0.99$ is a good option that balance the difference and convergence time. The curve for $\alpha=0.99$ is also consistent with the analytical one in Fig. \ref{fig-ratio}.

\begin{figure}[t]
\centering
\includegraphics[width=\figsize]{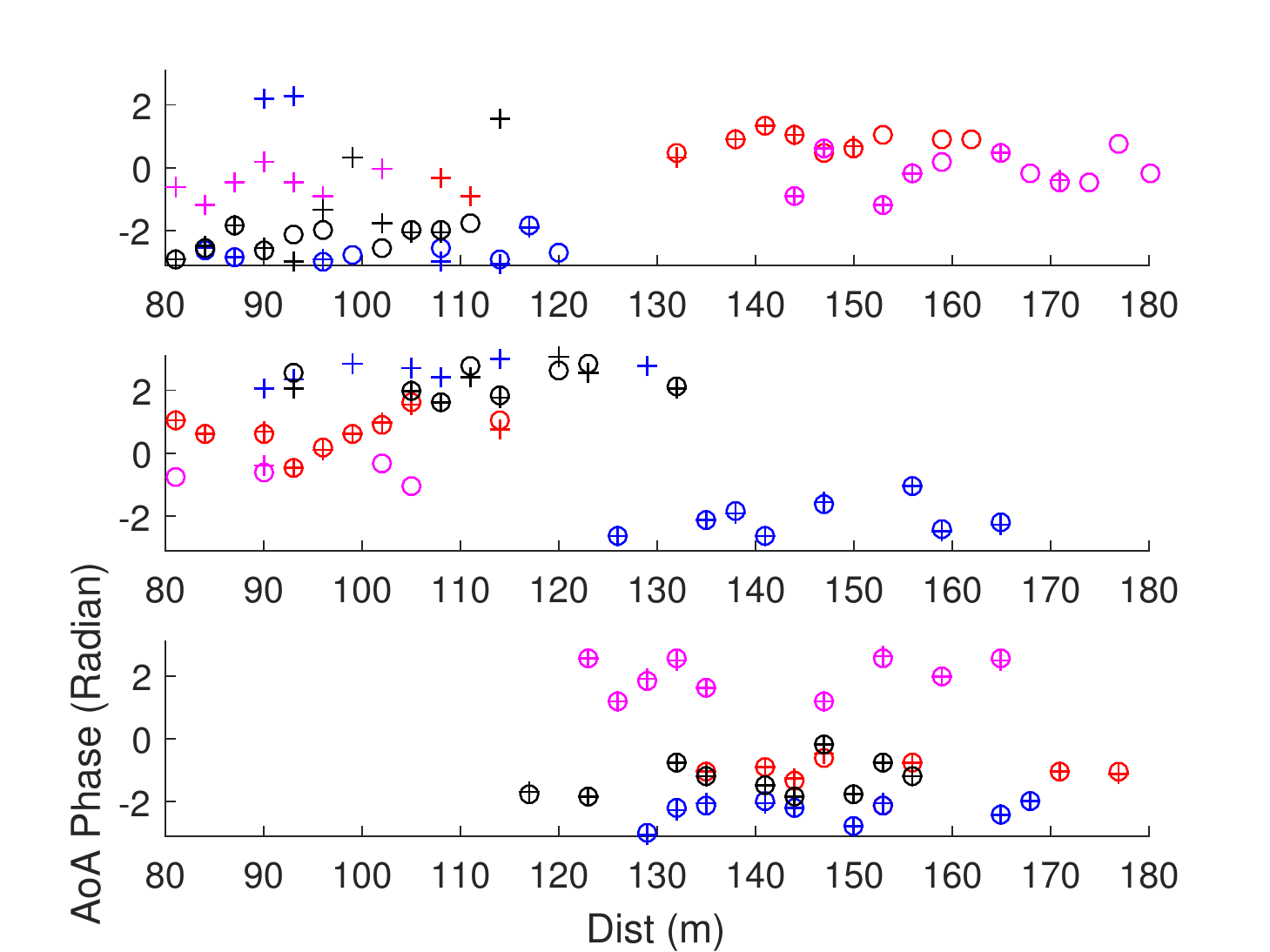}
\caption{Estimated and true sensing parameters AoA and distance obtained using the indirect method after clutter suppression. From top to bottom, $p=25, 50, 150$, respectively. Channel estimation $\eta=15$ dB.  }
\label{fig-est_sup}
\end{figure}

In Fig. \ref{fig-est_sup}, we show three random implementations with different values of $p$ used in estimating the clutter. From the top figure we can see both missed estimation for the current dynamic multipath and the estimate for the clutter and some residual dynamic multipath, which still have a strong presence in the subtracted signal. The middle one shows improved performance, and the bottom one achieves excellent estimation with clutter completely removed. This figure demonstrates the effectiveness of the proposed background subtraction algorithm.

\section{Conclusions}\label{sec-conclusion}
{
We developed a framework for a perceptive mobile network which integrates radio communication and sensing into one system, transforming the current communication-only mobile network. We presented a unified platform that enables both uplink and downlink sensing, using the uplink and downlink communication signals, respectively. We presented the required system modifications to enable this integration and formulated the mathematical model for sensing. We proposed the direct and indirect schemes based on 1D CS for estimating the sensing parameters, and the background subtraction method for clutter suppression. Our scheme is shown to work efficiently and is particularly suitable for the cases when sufficient measurements are only available in one domain. The perceptive mobile network can potentially facilitate many new sensing applications in smart city, smart home, smart car and transportation, while providing communication services. Although there are significant challenges and a long way ahead to make the perceptive mobile network fully operational, our work here is a solid first step, demonstrating the feasibility and providing a way to proceed. 
}

\bibliographystyle{IEEEtran}

\end{document}